\documentclass[twocolumn]{autart} %\documentclass[journal,twoside,web]{ieeecolor}
\usepackage{hyperref}
\usepackage{cite}
\usepackage{mathptmx} % assumes new font selection scheme installed
\usepackage{amsmath}
\usepackage{amssymb,amsfonts}
\usepackage{algorithmic}
\usepackage{graphicx}
\usepackage{textcomp}
\usepackage{xcolor}
\usepackage[scr]{rsfso}
\usepackage{soul}

 \newtheorem{lemmma}{Lemma}
\begin{document}
\begin{frontmatter}
\title{Adaptive Neural-Operator Backstepping Control \\ of a Benchmark Hyperbolic PDE}
\author%[author1]
{Maxence Lamarque}$^\text{a}$\ead{maxence.lamarque@etu.minesparis.psl.eu   },\ \ \
\author%[author2]
{Luke Bhan}$^\text{b}$\ead{lbhan@ucsd.edu},  \ \ \
\author%[author3]
{Yuanyuan Shi}$^\text{b}$\ead{yus047@ucsd.edu}, \ \ \ 
\author%[author4]
{Miroslav Krstic}$^\text{b}$\ead{krstic@ucsd.edu}\ \ \ 
\address%[author1,author2,author3,author4]
{$^\text{a}$\'{E}cole des Mines
de Paris, France,}
\address {$^\text{b}$ University of California, San Diego, USA}

\begin{abstract}
    To stabilize PDEs, feedback controllers require gain kernel functions, which are themselves governed by PDEs. Furthermore, these gain-kernel PDEs depend on the PDE plants' functional coefficients. The functional coefficients in PDE plants are often unknown. This requires an adaptive approach to PDE control, i.e., an estimation of the plant coefficients conducted concurrently with control, where a separate PDE for the gain kernel must be solved at each timestep upon the update in the plant coefficient function estimate. Solving a PDE at each timestep is computationally expensive and a barrier to the implementation of real-time adaptive control of PDEs. 
    
    Recently, results in neural operator (NO) approximations of functional mappings have been introduced into PDE control, for replacing the computation of the gain kernel with a neural network that is trained, once offline, and reused in real-time for rapid solution of the PDEs.     In this paper, we present the first result on applying NOs in adaptive PDE control, presented for a benchmark 1-D hyperbolic PDE with recirculation. We establish global stabilization via  Lyapunov analysis, in the plant and parameter error states, and also present an alternative approach, via passive identifiers, which avoids the strong assumptions on kernel differentiability. We then present numerical simulations demonstrating stability and observe speedups up to three orders of magnitude, highlighting the real-time efficacy of neural operators in adaptive control. Our code (\href{https://github.com/lukebhan/NeuralOperatorAdaptiveControl}{Github}) is made publicly available for future researchers.  

 \centering   {\sf\color{blue}This work was the subject of the 2023 Bode Prize Lecture by the last coauthor.} 
\end{abstract}

\end{frontmatter}
\allowdisplaybreaks
\setlength{\parskip}{.5em}  
\section{Introduction}

Following several papers in which PDE backstepping controllers were shown {\em robust} to the implementation of the gain kernels by replacing the solution of kernel PDEs by an offline-computed neural operator (NO) approximation of the kernel \cite{bhan_neural_2023,krstic2023neural,qi2023neural,wang2023deep,zhang2023neural,wang2023neural}, in this paper we introduce the {\em first adaptive} backstepping controller where the gain kernels are computed via NOs in real time, from online parameter estimates. We do so for a hyperbolic PDE with linear recirculation, the most accessible but nevertheless nontrivial (unstable) PDE system, with a functional coefficient that is unknown, and with boundary actuation. 

%As in \cite{bhan_neural_2023}, this approach is inherently model-based, taking the form of traditional adaptive PDE backstepping with approximation of the kernel operator via neural networks. Therefore, we reuse the exact 
We employ an (indirect) adaptive version of a standard PDE backstepping controller for a 1-D hyperbolic PDEs but with the analytical gain kernel replaced with the operator approximated equivalent. We then show, under the kernel operator approximation, global stability of the resulting closed-loop system via Lyapunov analysis and neural operator approximation theorems \cite{lu2021advectionDeepONet}, \cite{lanthaler2023nonlocal}. Furthermore, we present an alternative approach based on passive identifiers simplifying the assumptions on the gain-kernel derivatives at a cost of an increased dynamic order of the parameter estimator. 

This is the first result in which {\em offline learning} and {\em online learning} are both employed, working in tandem. Hence, it is important to explain these two distinct learning tasks. The operator from the plant coefficient to the kernel is learned offline --- once and for all. The unknown plant coefficient is learned online, continually, using a parameter estimator. The offline and online learners are combined through the adaptive gain, where the NO is evaluated, at each time step, for the new plant coefficient estimate. The NO speeds up the evaluation of the adaptive gain by about $10^3 \times$, relative to the hypothetical online solving of the gain kernel equation, and thus enables the real-time adaptive control of the PDE.

%We emphasize that this is the first work utilizing a NO, \textit{trained once, offline}, in conjunction with estimation of the PDE plant coefficients. As such, the power of the operator approximation is fully realized as \textit{at each timestep, the kernel needs to be recomputed according to the updated coefficient estimate}. Thus, the resulting methodology achieves numerical speedups on the magnitude of $10^3\times$ compared to traditional PDE solvers \textit{while retaining provable global stability guarantees}. 

Given the value of the $10^3 \times$ speedup in computing the adaptive gain, the code for all the computational tasks performed in relation to this adaptive design are made publicly available on \href{https://github.com/lukebhan/NeuralOperatorAdaptiveControl}{Github}. 
\vspace{-10pt}
\paragraph*{Stabilization of PDEs using backstepping-based adaptive control.}
The first investigations into backstepping-based adaptive control of PDEs were introduced for reaction-diffusion PDEs. Initially, a set of three approaches extending the simpler ODE counterparts were introduced: a  Lyapunov approach \cite{4623267}, a passive identification approach where one constructs an observer-like PDE system to estimate the plant parameter \cite{SMYSHLYAEV20071543}, and a swapping identifier where filters are introduced for the measurement to create a prediction error which can be minimized via standard techniques such as gradient descent \cite{SMYSHLYAEV20071557}. Papers \cite{BRESCHPIETRI20092074}, \cite{BRESCHPIETRI20141407} then extended these techniques to  adaptive control for systems with unknown delays in ODEs, and to wave PDEs. This paved the way for a swapping-based output-feedback extension to a single hyperbolic PDE \cite{BERNARD20142692}, and then to extensions to systems of hyperbolic PDES \cite{ANFINSEN201869}, \cite{anfinsen2016}, \cite{Anfinsen2019Adaptive}. Further, \cite{ZhuKrstic+2020} expanded the direction introduced by \cite{BRESCHPIETRI20092074} into a series of works on adaptive control of delay-systems. Concurrently, many works explored adaptive backstepping for different systems including coupled hyperbolic PDEs in \cite{7963000}, coupled hyperbolic PDE-PDE-ODE systems in \cite{9656694}, and the wave equation in \cite{WANG2020108640}. Lastly, we briefly mention the more recent works in adaptive control expanding into distributed input systems with unknown delays \cite{9528941} and event-triggered adaptive control of coupled hyperbolic PDEs \cite{WANG2021109637}, \cite{9735290}, \cite{KARAFYLLIS2019166}. 
\vspace{-10pt}

\paragraph*{Neural operator approximations for model-based PDE control.}
In a series of breakthrough innovations in the mathematics of machine learning \cite{lanthaler2023nonlocal}, \cite{Lu2021DeepONet}, \cite{lanthaler2022errorDeeponet}, universal operator approximation theorems have been developed which demonstrate that neural networks can effectively approximate mappings across function spaces. Naturally, the control community then capitalized on these results to approximate the kernel operator in PDE Backstepping. The first study in this direction was conducted for a 1D transport PDE in \cite{bhan_neural_2023}, and then later extended to both a reaction-diffusion PDE and observers in \cite{krstic2023neural}. In both works, the stability of the PDE under the approximated kernel is rigorously proved by employing the universal operator approximation theorem \cite{lu2021advectionDeepONet}. Following \cite{bhan_neural_2023}, \cite{krstic2023neural}, there have been a series of extensions where \cite{qi2023neural}, \cite{wang2023deep} developed similar results for hyperbolic and parabolic PDEs with delays. Paper \cite{zhang2023neural} then tackles the first application of NO approximations controlling the Aw-Rascale-Zhang(ARZ) PDE consisting of a set of second-order coupled hyperbolic PDEs describing traffic flows. Furthermore, \cite{wang2023neural} then considers NOs for a more general form of $2 \times 2$ hyperbolic PDEs with applications to oil drilling and shallow water wave modeling. Lastly, \cite{GS-preprint} employs neural operators for gain-scheduling of hyperbolic PDEs with nonlinear recirculation --- the first of such work where the kernel is recomputed at every timestep thus enabling real-time control of nonlinear PDEs. 

\vspace{-10pt}

\paragraph*{Contributions.} Two major advances in methodology and analysis are made. For Lyapunov-based and observer-based (passive) designs of update laws, two distinct neural operators are employed. For the Lyapunov update, a smoother NO is trained (the so-called ``full-kernel'' NO), leading to a target system with a homogeneous boundary condition and perturbations in the domain, whereas for the observer-based update, introduced in \cite{Anfinsen2019Adaptive}, a simpler but less smooth NO is trained (the so-called ``gain-only'' NO), eliminating the perturbation in the PDE's domain but making the boundary condition perturbed. These two designs give rise to distinct mathematical issues to overcome. The paper not only solves the technical problems that arise in NO-based adaptive PDE control but also illuminates the tradeoff between the two NO approaches. 

The key novel mathematical challenge overcome in this paper, relative to the papers \cite{bhan_neural_2023,krstic2023neural,qi2023neural,wang2023deep,zhang2023neural,wang2023neural} in which the robustness to NO approximating of the gain is established, is that the updating of the plant coefficient, and the associated updating of the kernel through the NO, gives rise to not only a potentially high rate of change in the adaptive gain but also a potentially high rate of change of the error in the NO approximation of the adaptive gain. This mathematical challenge is handled differently in the Lyapunov/full-kernel and observer-based/gain-only approaches. Each approach has its merit and each of the proof procedures has an educational value to the reader aspiring to pursue extensions of NO-enabled adaptive control of PDEs.

The most obvious contribution is in the enablement of real-time adaptive PDE control, through a $10^3\times$ speedup in the computation of the adaptive gain. 
\vspace{-10pt}

\paragraph*{Paper outline.}
In Sec. \ref{sec-exact-adap}, we briefly restate the unpublished but relatively easy result for adaptive PDE backstepping of hyperbolic PDEs with recirculation. In Sec. \ref{section_kernel_ppties}, we prove both existence and boundedness of the exact backstepping kernel and its derivative. In Sec. \ref{section:no_ppties}, we then present the neural operator approximation theorem and show the adaptive backstepping kernel can be approximated by a neural operator. Next, in Sec. \ref{sec-Lyap-design}, we  give the paper's main result presenting stability of the closed loop feedback system under the neural operator. We follow the result with a proof in Sec. \ref{section:lyap_computations} via Lyapunov analysis. In Sec. \ref{section:modular_design_with_passive_identifier}, we present an alternative approach, via a modular design with a passive identifier that avoids the approximation of the kernel's derivative and thus the strong assumptions about the kernel's differentiablity required for Lyapunov analysis. Lastly, in Sec. \ref{section:simulations}, we present numerical simulations highlighting the theoretical stability result and calculate the numerical speedups gained from the neural operator approximation.

\vspace{-10pt}
\paragraph*{Notation.}
\begin{table}[]
    \centering
    \begin{tabular}{|l|c|}
    \hline
        exact operator & $\mathcal K$ 
\\ \hline 
        neural (approximate) operator & $\hat{\mathcal K}$
\\ \hline\hline
exact kernel & $k=\mathcal K (\beta)$
\\ \hline
exact estimated kernel & $\breve k = \mathcal K(\hat \beta)$
\\ \hline
approximate estimated kernel & \\
(adaptive kernel) & 
$\hat k = \hat{\mathcal K}(\hat \beta)$ 
\\ \hline
    \end{tabular}
    \caption{Nomenclature for offline and online kernel learning}
    \label{tab:nomenclature}
\end{table}
We present the nomenclature for the offline and online kernel in Table \ref{tab:nomenclature}. We abbreviate the partial derivative as
$      a_{x}(x, \nu) = \frac{\partial a}{\partial x } (x, \nu)$.
    For a function $a$ defined on $[0, 1] \times \mathbb{R}^+$ we denote the spatial $L^2$ norm as 
$    \|a(t)\| = \sqrt{\int_{0}^{1} a^2(x, t)dx}$, which is a function of $ t \in \mathbb{R}^+$. 
  %   \|a\|&:& \mathbb{R}^+ \to \mathbb{R}^+ \nonumber \\
  %        &&t \to \|a(t)\|
  % \end{eqnarray}
%  We denote the set $\mathcal{T}$ as  $\mathcal{T} := \left \{ (x, y) | \, 0 \leq y \leq x \leq 1 \right \}$.
We denote the convolution operation (which is commutative) by
    \begin{equation}
     a * b (x) = b*a  (x) = \int_{0}^{x} a(x-y) b(y) dy\,.
    \end{equation}
%and note that it is commutative, $    b*a = a*b$. 

\setlength{\parskip}{1em}  

\section{Exact Adaptative  PDE Backstepping for a Hyperbolic PDE with Recirculation}
\label{sec-exact-adap}

% The transformation maps the system \eqref{eq:vt_def}, \eqref{eq:v1_def} to 
 We consider the following hyperbolic PDE---transport PDE with recirculation,
\begin{eqnarray}
    u_t(x, t) &=& u_x(x, t) + \beta(x)u(0, t)\,, \quad \forall (x, t) \in [0, 1) \times \mathbb{R}^+ \label{eq:ut_def} \\
    u(1, t) &=& U(t) \label{eq:u1_def} \,,
%    \\ \beta(x) &:=& q(x, 0) + g(x) - \int_{0}^{x} q(x, y) g(y) dy
  \end{eqnarray}
%  which we want to stabilize at the equilibrium $u \equiv 0$, 
  where $\beta$ is an unknown function
  %upper bounded by $B := G+ Fe^F (G+1)$ and 
  to be estimated online using an estimate $\hat{\beta}(x, t)$. % of $\beta$. 
  
We employ an {\em adaptive backstepping transformation} given by
  \begin{eqnarray}
  \label{eq:w_def} 
    w(x, t) &=& u(x, t) - \breve{k} * u (x, t)\,,
%&&
 \qquad           \forall (x, t) \in [0, 1] \times \mathbb{R}\,,
  \end{eqnarray}
  where $\breve k$ is the (online) backstepping kernel solution of the Volterra equation
  \begin{equation}
    \breve{k}(x, t) = -\hat{\beta}(x, t)  + \hat{\beta} * \breve{k} (x, t)\,, \quad (x, t) \in [0, 1] \times \mathbb{R}^+\,.
    \label{eq-exact-adapt-Volterra-eqn}
  \end{equation}

  The transformation \eqref{eq:w_def} maps the system \eqref{eq:ut_def}, \eqref{eq:u1_def} to the perturbed target system
  \begin{eqnarray}
    \label{eq:w1_def}
    w_t(x, t) &=& w_x(x, t) \nonumber \\ &&  + \left [\tilde{\beta}(x, t) -  \breve{k} * \tilde{\beta}(x, t)\right ] w(0, t) - \Omega(x, t) \label{eq:wt_def}\,, \\
    w(1, t) &=& 0, 
    %\quad \forall(x, t) \in [0, 1] \times \mathbb{R}^+ 
  \end{eqnarray}
  where
  \begin{eqnarray}
    \tilde{\beta}(x, t) &=& \beta(x, t) - \hat{\beta}(x, t),  \label{eq:tilde_beta_def}\\
    \Omega(x, t) &=& \breve{k}_t * (w -\breve l * w) (x, t) \,,
%    = \breve{k}_t * \left(1-\mathcal{K}\left(\breve k\right)\right)*)w (x, t) 
    \label{eq:omega_def} \\
% l(x, t) &=& \breve{k}(x, t) + \breve{k} * l (x, t)
%     = - \mathcal{K}\left(\breve k\right)(x,t)
%     = - \mathcal{K}\circ \mathcal{K}\left(\hat\beta\right)(x,t)
\breve l &=& -\breve{k} + \breve{k} * \breve l 
    =  \mathcal{K}\left(\breve k\right)
    =  \mathcal{K}\circ \mathcal{K}\left(\hat\beta\right)
    =\hat\beta\,.
%\\ \nonumber &&  \qquad\qquad\qquad \forall (x, t) \in [0, 1] \times \mathbb{R} \label{eq:l_def}.
  \end{eqnarray}
  Note that the boundary condition \eqref{eq:w1_def} gives from \eqref{eq:w_def} the feedback law 
  \begin{equation}
    U(t) = \breve{k} * u (1, t)
    = \mathcal{K}(\hat\beta)*u(1,t).
    \label{eq:U_def_true_k}
  \end{equation}

  We first state an adaptive control design for the adaptative problem with the exact backstepping kernel $\breve{k}$. The next theorem serves only as a guidance for what we seek to achieve under a {\em NO-based approximate adaptive backstepping} design. We omit the theorem's proof since it can be deduced from the proof of our main result in Theorem \ref{thm-main}.

\bigskip
\begin{thm}{\em [Full-state {\em exact} adaptative control design.]}
Consider the plant \eqref{eq:ut_def}-\eqref{eq:u1_def} in feedback with the control law
  \begin{eqnarray}
  \label{eq-exact-adapt-bkst-control}
    U(t) = \int_{0}^{1} \breve{k}(1-y, t)u(y, t)dy, \quad t \geq 0
  \end{eqnarray}
  where $\breve{k}$ is solution of the Volterra integral equation \eqref{eq-exact-adapt-Volterra-eqn}.
  % \begin{equation}
  % \label{eq-exact-adapt-Volterra-eqn}
  %   \breve{k}(x, t) = -\hat{\beta}(x, t) + \hat{\beta} * \breve{k} (x, t),  \quad (x, t) \in [0, 1] \times \mathbb{R}^+.
  % \end{equation}
For all $c > 0$ and all $B>0$ such that $\|\beta\|_{\infty} \leq B$, there exists $\gamma^*(c, B) = \mathcal{O}_{c \to \infty}(e^{-c}) > 0$ with a decreasing dependence on $B$, such that for all $ \gamma \in (0, \gamma^*)$, any intial condition $\hat{\beta}(\cdot, 0) \in %\mathcal{C}^1([0, 1],
{ \mathcal{C}^0([0, 1],}
\mathbb{R})$ satisfying $\|\hat{\beta}(\cdot, 0)\|_{\infty} \leq B$,
 the update law 
\begin{eqnarray}
   \label{eq-update-betahat-exact}
  \hat{\beta}_t (x, t) &:=&  {\rm Proj} (\tau(x, t), \hat{\beta}(x,t))
   ,\quad\forall (x, t) \in [0, 1] \times \mathbb{R}^+\,, \\ 
\tau(x, t) &:=& \nonumber \frac{\gamma}{1 + \|w(t)\|^2_c} \bigg[e^{cx} w(x, t) - \\ && \int_{x}^{1}  \breve{k}(y-x, t)e^{cy} w(y, t) dy \bigg] w(0, t), 
   \label{eq-tau-exact}
\end{eqnarray}
  where
  \begin{eqnarray}
    w &=& u - \breve{k} * u\,,  \\
    \|w(t)\|^2_c &=& \int_{0}^{1} e^{cx} w^2(x, t) dx\,, 
  \end{eqnarray}
with the projection operator $ {\rm Proj}: \mathbb{R} \times [0, B] \to \mathbb{R} $ defined as\footnote{The projector operator defined here is not continuous. Hence, the solutions of the PDE system are in the Fillipov sense. To avoid the discontinuity, one would add a boundary layer of width $\delta > 0$. But to avoid having the exposition drifting into inessential technicalities, we use the common discontinuous projection  \eqref{eq:proj_def}. 
}
\begin{eqnarray}
            %(a, b) \to 
            {\rm Proj}(a, b) := 
            \begin{cases}
              0, &\mbox{ if } |b| = B \mbox{ and } ab > 0 \\
              a, &\mbox{else }
            \end{cases}
  \label{eq:proj_def}
\end{eqnarray}
guarantees that 
\begin{eqnarray}
  \Gamma(t)&\leq& R(e^{\rho \Gamma(0)} - 1)\,, \qquad \forall t\geq 0, 
    \\
    \Gamma(t) &=& \int_0^1 \left[u^2(x,t) + \left(\beta(x) - \hat\beta(x,t)\right)^2\right] dx
\end{eqnarray}
for constants $\rho, K  > 0$ and, in addition, 
%  $u$ , w$ uniformly converges to $\tilde{0}$ as $t \to \infty$  and $\|u(t)\| %, \|w(t)\|
  $u(x,t)  \underset{t \to \infty}{\to} 0$ for all $x\in[0,1]$.
\end{thm}

In summary, with the {\em exact} adaptive backstepping feedback law \eqref{eq-exact-adapt-bkst-control}, \eqref{eq-exact-adapt-Volterra-eqn}, \eqref{eq-update-betahat-exact}, \eqref{eq-tau-exact}, the equilibrium $(u(x),\hat\beta(x)) \equiv (0, \beta(x))$ is globally stable in the $L^2$ sense and the state $u(x,t)$ is regulated to zero pointwise in $x$. The computationally intensive part of implementing this feedback law is that the Volterra equation \eqref{eq-exact-adapt-Volterra-eqn} needs to be solved (in $x$) at each time ``step'' $t$. It is for this reason that we seek a neural operator approximation $\hat{\mathcal{K}}: \hat\beta \mapsto \hat k$ to the exact adaptive backstepping gain operator $\mathcal{K}: \hat\beta \mapsto \breve k$, which would require only a neural network {\em evaluation} at each $t$, rather than a solution to a Volterra equation.

%With this feedback law, provided that we choose a correct update law for $\hat{\beta}_t$, we can achieve global assymptotically stability for the equilibrium point $u \equiv 0$. But computing this feedback law requires online computations and numerically solving the volterra equation \eqref{eq-exact-adapt-Volterra-eqn} at each time $t$ which is costly. That is why this document aims at achieving a similar result with an approximated estimated backstepping kernel $\hat k$ provided by a neural operator.

  % In Section~\ref{section_kernel_ppties} we first establish the properties of the direct and inverse backstepping kernels. Then, in Section~\ref{section:no_ppties}, we first introduce the neural operator used to approximate the map for the parameter update $\hat\beta$ to the control gain $\hat k$. In Section \ref{sec-Lyap-design} we stating the main result, stabilization with a DeepONet implementation of the adaptive gain. Finally, in  Section~\ref{section:lyap_computations} we perform a Lyapunov analysis to prove the main result stated in Section \ref{sec-Lyap-design}.

\section{Backstepping Kernel Properties}
\label{section_kernel_ppties}

This section introduces results on the exact adaptive backstepping kernel $\breve k$ in \eqref{eq-exact-adapt-Volterra-eqn}.

%\bigskip
\begin{lemmma}\label{lemma_k_existence_and_bound}
{\em [Existence and upper bound for kernel and its derivative]}
Let $B > 0, \hat{\beta} \in \mathcal{C}^0([0, 1] \times \mathbb{R}^+, \mathbb{R})$ such that $\|\hat{\beta}\|_{\infty} \leq B$ and consider 
 the Volterra equation \eqref{eq-exact-adapt-Volterra-eqn}, reiterated here for convenience, 
\begin{eqnarray}
\label{eq:kernel_lemma_def}
  \breve{k} (x, t) -  \hat{\beta} * \breve{k} (x, t)+\hat{\beta}(x, t)=0\,, \quad (x, t) \in [0, 1] \times \mathbb{R}^+. 
\end{eqnarray}
There exist a unique $\mathcal{C}^0 ([0, 1] \times \mathbb{R}^+, \mathbb{R})$ solution $\breve{k}$ that satisfies 
\begin{equation}
  \|\breve{k}\|_{\infty} \leq Be^B.
  \label{eq:kernel_bound}
\end{equation}
If, in addition, 
%we also make the assumption that 
$\hat{\beta}_t$ exists and is continuous with respect to $x$ on $[0, 1] \times \mathbb{R}^+$ such that
\begin{eqnarray}
  \|\hat{\beta}_t\|_{\infty, [0, 1] \times [0, T]} < \infty, \quad \forall T > 0, \label{eq:beta_hat_t_bound}
\end{eqnarray}
then $\breve{k}_t$ exists, is continuous with respect to $x$ on $[0, 1] \times \mathbb{R}^+$, and satisfies
\begin{eqnarray}
  \|\breve{k}_t(t)\| &\leq& \|\hat{\beta}_t(t)\|(1 + Be^B(2 + Be^B)), \quad t \geq 0 \label{eq:k_l2_bound}\,.
\end{eqnarray}
\end{lemmma}

\begin{pf}
  Let $B > 0$, $\hat{\beta} \in \mathcal{C}^0([0, 1] \times \mathbb{R}^+, \mathbb{R})$ such that $\|\hat{\beta}\|_{\infty} \leq B$. We notice that \eqref{eq:kernel_lemma_def} is just a Volterra integral equation since $\hat{\beta}$ is continuous. The existence and continuity of $\breve{k}$ follows. Also, note that \eqref{eq:kernel_lemma_def}
  \begin{eqnarray}
    |\breve{k}(x, t)| \leq B + \int_{0}^{x} |\hat{\beta}(x-y, t)|.|\breve{k}(y, t)| dy. 
  \end{eqnarray}
  Then Grönwall's lemma gives \eqref{eq:kernel_bound}. 
  We now prove the existence and continuity with respect to $x$ of $\breve{k}_t$ on $[0, 1] \times \mathbb{R}^+$. To do so we use a successive approximation approach. We introduce the sequence
  \begin{eqnarray}
    \Delta k^0 &:=& -\hat{\beta} \label{eq:Delta_k_0_def}\,, \\
    \Delta k^{n+1} &:=& \hat{\beta} * \Delta k^n \label{eq:delta_k_n+1_def}\,.
  \end{eqnarray}
  Through iteration we have
  \begin{eqnarray}
    |\Delta k^n(x, t)| &\leq& \frac{B^{n+1} x^n}{n!}, \quad (x, t) \in [0, 1] \times \mathbb{R}^+\,, \label{eq:kn_bound}
  \end{eqnarray}
  From which we have
  \begin{equation}
    \breve{k} = \sum_{n=0}^{\infty} \Delta k^n \label{eq:k_sum}\,.
  \end{equation}
  We will be proving that the series $\sum_{n=0}^{\infty} \Delta k^n_t$ uniformly converges on each compact $[0, 1] \times [0, T]$, $ T > 0$. We begin by introducing the function
  \begin{equation}
    B_t(T) := \|\hat{\beta}_t\|_{\infty, [0, 1] \times [0, T]} < \infty, \quad T > 0\,, 
    \label{eq:beta_hat_bound}
  \end{equation}
  with assumption \eqref{eq:beta_hat_bound}.
  We can prove through induction that $\forall n \in \mathbb{N}$, $\Delta k^n_t$ exists, is continuous with respect to $x$, and satisfies
  \begin{eqnarray}
    |\Delta k^n_t(x, t) | &\leq& \frac{(n+1) \alpha(T) x^n}{n!}, \quad \forall (x, t) \in [0, 1] \times [0, T],\label{eq:delta_kn_t_bound} \\
    \alpha(T) &:=& {\rm max} \left \{B_t(T), B \right \}.
  \end{eqnarray}
  To do so we have to take the derivative of \eqref{eq:delta_k_n+1_def} with respect to $t$ using the Leibniz theorem for the derivation of integral with parameter. Notice that we use the `strong' version of this theorem that doesn't require $\hat{\beta}_t$ to be continuous with respect to $t$.
  From \eqref{eq:delta_kn_t_bound}, we have that the series $\sum_{n=0}^{\infty} \Delta k^n_t$ uniformly converges on $[0,1] \times [0, T]$, for all $T > 0$. We thus have the existence and continuity with respect to $x$ of $\breve{k}_t$ on $[0, 1] \times \mathbb{R}^+$. We can then take the derivative of \eqref{eq:kernel_lemma_def} with respect to $t$, which gives the following inequality satisfied by $\breve{k}_t$, 
  \begin{eqnarray}
    |\breve{k}_t(x, t)| &\leq& |\hat{\beta}_t(x, t)| + Be^B \|\hat{\beta}_t(x, t)\| \nonumber \\ && + \int_{0}^{x} |\hat{\beta}(x-y, t)| |\breve{k}_t(y, t)| dy, \label{eq:kt_inequality} \\
                        &&\forall (x, t) \in [0, 1] \times \mathbb{R}^+ \nonumber\,.
  \end{eqnarray}
  Then using Grönwall lemma on \eqref{eq:kt_inequality}, we arrive at
    \begin{eqnarray}
        |\breve{k}_t(t, x)| &\leq& |\hat{\beta}_t(x, t)| + Be^B \|\hat{\beta}_t(t)\| (2 + Be^B), \nonumber \\ &&  \forall (x, t) \in [0, 1] \times \mathbb{R}^+\,.  \label{eq:kt_bound}
    \end{eqnarray}
    From \eqref{eq:kt_bound}, using the triangular inequality we have \eqref{eq:k_l2_bound}.
\end{pf}

\section{Neural Operator Approximation of Backstepping Kernel}
 \label{section:no_ppties}
 
 Explicitly solving the Volterra equation \eqref{eq-exact-adapt-Volterra-eqn} satsfied by $\breve k$ is almost never feasible, and solving it numerically is  expensive. We design an approximate operator which, for an estimate $\hat\beta$ of the unknown $\beta$ in the plant  $\eqref{eq:ut_def}$ produces an approximate adaptive kernel $\hat{k}$, generated by evaluating a neural operator for the input $\hat\beta$. 
 
 For such an approach to guarantee stabilization when the exact adaptive kernal $\breve k$ is replaced by the approximate kernel $\hat k$, we need to design a neural operator that keeps the approximation error $\breve k - \hat k$ small in a suitable sense. To produce such a neural operator, we recall the DeepONet universal approximation theorem. 

\begin{thm}{\em [DeepOnet universal approximation theorem \cite{lu2021advectionDeepONet}.]}
\label{theorem:DeepOnet}
    Let $X \subset \mathbb{R}^{d_x}$ and $Y \subset \mathbb{R}^{d_y}$ be compact sets of vectors $x \in X$ and $y \in Y$, respectively. Let $\mathcal{U}: X \to U \subset \mathbb{R}^{d_u}$ and $\mathcal{K}: Y \to V \subset \mathbb{R}^{d_v}$ be sets of continuous functions $u(x)$ and $v(y)$, respectively. Let $\mathcal{U}$ also be compact. Assume the operator $\mathcal{G}: U \to V$ is continuous. Then, for all $\epsilon > 0$, there exist $m^*, p^* \in \mathbb{N}$ such that for each $m \geq m^*$, $p \geq p^*$, there exist $\theta^{(k)}$, $v^{(k)}$, neural networks $f_N(\cdot; \theta^{(k)})$, $g_N(\cdot; v^{(k)})$, $k = 1, \ldots, p$, and $x_j \in X$, $j = 1, \ldots, m$, with corresponding $\mathbf{u}_m = (u(x_1), u(x_2), \ldots, u(x_m))^T$, such that
    \begin{equation}
      |\mathcal{G}(u)(y) - \mathcal{G}_{\mathbb{N}}(\mathbf{u}_m)(y)| < \epsilon
    \end{equation}
    for all functions $u \in \mathcal{U}$ and all values $y \in Y$ of $\mathcal{G}(u)$.
    Where 
    \begin{equation}
      \mathcal{G}_{\mathbb{N}}(y) = \sum_{k=1}^p g^{\mathcal{N}}(\mathbf{u}_m; v^{(k)})f^{\mathcal{N}}(y; \theta^{(k)})
    \end{equation}
\end{thm}

We denote by $\mathcal{K}: H \to \mathcal{C}^0([0, 1],  \mathbb{R})$ the operator %defined as
%that we are trying to learn
\begin{equation}
  \label{eq:op_def}
  %\mathcal{\breve{K}} = 
  \mathcal{K}: \, \hat{\beta}(\cdot, t) \mapsto \breve k(\cdot,t) 
\,,
%=: \mathcal{K}\left(\hat{\beta}(\cdot, t)\right) , 
%  \mathcal{K}: \hat{\beta}(\cdot, t) \in H \to \breve k(\hat{\beta}(\cdot, t)) \in \mathcal{C}^0([0, 1],  \mathbb{R})
\end{equation}
where $\breve k(\cdot, t)$ is the solution to the Volterra integral equation \eqref{eq-exact-adapt-Volterra-eqn} at a specific time $t \geq 0$. Since the parameter estimate $\hat\beta(x,t)$ is time varying, its time derivative affects the closed-loop system. For this reason, it is not enough to approximate only $\breve{k}$. Its derivative $\breve{k}_t(\cdot, t)$ also must be approximated, at each time $t$. It is crucial to note that, while we are concerned about approximating a derivative in time, $\breve{k}_t(x, t)$, it is only an accurate approximation of this quantity as a function of $x$ that is needed. 
%To be specific, $\partial_t \left(\breve k(x,t) - \hat k(x,t)\right)$ need not be small; it is only $\breve k_t(x,t) - \hat k_t(x,t)$ that we insist on being small. 

For this purpose we denote by $\mathcal{M}: H^2 \to \mathcal{C}^0([0, 1],\mathbb{R})^2$  the operator
%, for which we then seek a neural operator approximation:
\begin{eqnarray}
  \mathcal{M}:\, 
%  H^2 \to \mathcal{C}^0([0, 1], \mathbb{R})^2\nonumber \\&& 
 (\hat{\beta}(\cdot, t), \hat{\beta}_t(\cdot, t))  \mapsto (\mathcal{K}(\hat{\beta}(\cdot, t)), \mathcal{K}_1(\hat{\beta}(\cdot, t), \hat{\beta}_t(\cdot, t))
\end{eqnarray}
where $\mathcal{K}_1$ is defined as the operator that maps $(\hat{\beta}(\cdot, t), \hat{\beta}_t(\cdot, t))$  into the solution $\breve{k}_1$ of the Volterra equation
\begin{eqnarray}
  &&\nonumber \breve{k}_1(x,t) - 
   \int_{0}^{x} \hat{\beta}(x-y, t) \breve{k}_1(y,t)dy
  \\ &&  +\hat{\beta}_t(x, t)   - \int_{0}^{x} \hat{\beta}_t(x-y, t) \breve{k}(y, t)dy =0\,,
  \quad 
  x \in [0, 1]\,,  
\end{eqnarray}
namely,
\begin{equation}
    \breve k_1 = \mathcal{K}_1\left(\hat\beta,\hat\beta_t\right) := \mathcal{B}\left( -\hat\beta_t + \hat\beta_t *\mathcal{K}\left(\hat\beta\right),\mathcal{K}\left(\hat\beta\right)\right)\,,
\end{equation} 
which is explicitly given by the expression in Lemma \ref{lem-k0k1}. The set $H$ denotes the subset of $\mathcal{C}^0([0, 1], \mathbb{R})$  endowed with the supremum ($\|\cdot\|_\infty$) norm, such that all $\alpha \in H$  satisfy
\begin{itemize}
  \item $\|\alpha\|_{\infty} \leq M$
  \item $\alpha$ is K-Lipschitz
\end{itemize}
where $M, K > 0$ can be as large as required. 

To approximate the operator $\mathcal{M} = \left(\mathcal{K}, \mathcal{K}_1\right)$ by a DeepONet, the conditions of Theorem \ref{theorem:DeepOnet}, require us to define a specific compact set of the $\hat{\beta}(\cdot, t), \hat{\beta}_t(\cdot, t)$ functions. This is the purpose of introducing the set $H$, which is compact by the Arzel\`{a}-Ascoli theorem, and so is the set  $H^2:= H\times H$. 

We have proven that the operator $\mathcal{M}$ is continuous (and even Lipschitz) in Lemma 2 of \cite{GS-preprint}. We can then state the following theorem, which is a consequence of Theorem \ref{theorem:DeepOnet}.

\begin{thm}\label{theorem:NO_k}
{\em [Existence of a neural operator approximating the kernel.]}
For all $\epsilon > 0$, there exist a neural operator $(\hat{\mathcal{K}}, \hat{\mathcal{K}}_1)$ such that for all $\hat{\beta}(\cdot, t), \hat{\beta}_t(\cdot, t) \in H$ and for all $\forall x \in [0, 1] $, 
\begin{align}
  &\left|\mathcal{K}\left(\hat{\beta}(\cdot, t)\right)(x) - \hat{\mathcal{K}}\left(\hat{\beta}(\cdot, t)\right)(x)\right| \nonumber \\
  +&\left|\mathcal{K}_1\left(\hat{\beta}(\cdot, t), \hat{\beta}_t(\cdot, t)\right)(x) - \hat{\mathcal{K}}_1\left(\hat{\beta}(\cdot, t), \hat{\beta}_t(\cdot, t)\right)(x)\right|< \epsilon 
    \,. \label{eq:khat_property} 
\end{align}
\end{thm}

\section{DeepONet-Approximated Lyapunov Adaptive PDE Backstepping Design}
\label{sec-Lyap-design}

The stabilizing property of the adaptive backstepping controller employing an approximate estimated kernel is given in the next theorem, our main result. 

\begin{thm}{\em [Stability of {\em approximate} Lyapunov adaptive backstepping control.]} 
\label{thm-main}
  For all $B, %B_L, 
  c > 0$, 
  %{\color{red} $B_L$ not defined }
  there exists $\epsilon_0(B, c) = \underset{c \to \infty}{\mathcal{O}}(ce^{\frac{-c}{2}}), \gamma_0(B, c) = \underset{c \to \infty}{\mathcal{O}}(e^{-c})$ with a decreasing dependence on the argument $B$ such that for all neural operator approximations $\hat{k}=\hat{\mathcal{K}}\left(\hat\beta\right)$ of accuracy $\epsilon \in (0, \epsilon_0)$ provided by Theorem \ref{theorem:NO_k}, all $\gamma \in (0, \gamma_0)$, and all $\beta, \hat{\beta}(\cdot, 0) 
%\in \mathcal{C}^1([0,1], \mathbb{R})
$ that are Lipschitz and satisfy $\|\beta\|_{\infty},\|\hat{\beta}(\cdot, 0)\|_{\infty} \leq B$,
%   \end{eqnarray}
  the feedback law
  \begin{equation} \label{eq-control}
    U(t) = \int_{0}^{1} \hat{k}(1-y, t) u(y, t)dy\,, 
  \end{equation}
  and the update law
\begin{eqnarray}
  \label{eq-update-betahat-approx}   
  \hat{\beta}_t (x, t) &:=&  {\rm Proj} (\tau(x, t), 
%  \tau_x(x, t), 
  \hat{\beta}(x,t), 
%  \hat{\beta}_x(x, t)
  )
   ,\quad\forall (x, t) \in [0, 1] \times \mathbb{R}^+\,, 
\\
\label{eq-tau-approx}
\tau(x, t) &:=& \frac{\gamma}{1 + \|w(t)\|^2_c} \bigg[e^{cx} w(x, t) \nonumber \\ && - \int_{x}^{1}  \hat{k}(y-x, t)e^{cy}w(y, t) dy \bigg] w(0, t), 
\end{eqnarray}
where
  \begin{eqnarray}
  \label{eq-bkst-w-khat}
    w &=& u - \hat{k} * u \,, \\ \label{eq-norm-w}
    \|w(t)\|^2_c &=& \int_{0}^{1} e^{cx} w(x, t) dx\,,
  \end{eqnarray}
%  with the projection operator \eqref{eq:proj_def}
%$ {\rm Proj}: \mathbb{R}^2 \times [0, B] \times [0, B_L] \to \mathbb{R} $ defined as
%\footnote{The projector operator defined here is not continuous, which requires to understand the solutions of the PDEs in the Fillipov sense. To avoid this discussion one could add a boundary layer of width $\delta > 0$ but to avoid having the exposition drift into inessential technicalities, we use the common discontinuous projection  \eqref{eq:proj_def}. }
% \begin{eqnarray}
%             %(a, b) \to 
%             {\rm Proj}(a, b) := 
%             \begin{cases}
%               0, &\mbox{ if } |b| = B \mbox{ and } ab > 0 \\
%               a, &\mbox{else }
%             \end{cases}
% \end{eqnarray}
% MAYBE REWORK THE PROJECTOR INTO
% \begin{eqnarray}
%             {\rm Proj}(a, a_x, b, b_x) := 
% \begin{cases}
%   0, &\mbox{ if } |b_x| = B_L \mbox{ and } a_x b_x > 0 \mbox{ or } |b| = B \mbox{ and } a b > 0  \\
%   a, &\mbox{else }
% \end{cases} \label{eq:weird_proj_def}
% \end{eqnarray}
guarantee that all solutions for which $\hat\beta_t(\cdot,t)$ remains in $H$, $\hat k(\cdot,t)$ is differentiable, and $\hat\beta(\cdot,t)$ remains Lipschitz
%, and \allowbreak $- \tilde\beta \, {\rm Proj} (\tau, \tau_x, \hat{\beta}, \hat{\beta}_x)(\cdot, t)  \leq - \tilde\beta \, \tau(\cdot,t)$
for all time satisfy
\begin{eqnarray}
\label{eq-globstab}
  \Gamma(t)&\leq& R(e^{\rho \Gamma(0)} - 1)\,, \qquad \forall t\geq 0, 
\\
\label{eq:gamma_def}
    \Gamma(t) &=& \int_0^1 \left[u^2(x,t) + \left(\beta(x) - \hat\beta(x,t)\right)^2\right] dx\,, 
\end{eqnarray}
for constants $\rho, R  > 0$ and, in addition, 
%  $u$ , w$ uniformly converges to $\tilde{0}$ as $t \to \infty$  and $\|u(t)\| %, \|w(t)\|
\begin{equation}
    u(x,t)  \underset{t \to \infty}{\to} 0\,, \qquad \forall x\in[0,1]\,.
\end{equation}
\end{thm}

The assumptions that $\hat\beta_t(\cdot,t)$ remains in $H$ and that $\hat k(\cdot,t)$ is differentiable for all time are  strong and not a priori verifiable. They arise from the fact that  in the Lyapunov design it is necessary to approximate the update rate $\hat k_t$ of the approximated kernel $\hat k$. This motivates us to pursue, in Section \ref{section:modular_design_with_passive_identifier}, an alternative {\em modular design with a passive identifier}, which doesn't require an approximation of the derivative of the approximated kernel and, hence, doesn't require these strong assumptions on $\hat\beta_t(\cdot,t)$ and $\hat k_t(\cdot,t)$.

% \begin{remark}\em
%   This theorem relies on the nontrivial assumptions
%   \begin{itemize}
%     \item For all $t \geq 0$, $\hat{\beta}_t \in H$
%     \item For all $t \geq 0$, we have that $\hat{k}$ is differentiable with respect to time $t$ and equals the quantity $\hat{k}_2$, which is the neural operator approximation of the solution to the Volterra integral equation
%       \begin{eqnarray}
%         \breve{k}_2 = -\hat{\beta}_t + \hat{\beta} * \breve{k}_2 + \hat{\beta}_t * \mathcal{K}(\hat{\beta})
%       \end{eqnarray}
%   \end{itemize}
%   This statement aims more at exposing the limitations with Lyapunov design with the DeepOnet approximation than to state a result. To overcome these two powerful assumptions, in Section \ref{section:modular_design_with_passive_identifier} we introduce  a modular design with a passive identifier, which doesn't require these assumptions.
% \end{remark}

Our parameter update law \eqref{eq-tau-approx} is a replica of %\eqref{eq-update-betahat-exact}, 
\eqref{eq-tau-exact}
% , namely, 
% \begin{eqnarray}
%   \label{eq-update-betahat-approx}   
%   \hat{\beta}_t (x, t) &:=&  {\rm Proj} (\tau(x, t), \hat{\beta}(x,t))
%   ,\quad\forall (x, t) \in [0, 1] \times \mathbb{R}^+
%   \\
%     \tau(x, t) &:=& \frac{\gamma}{1 + \|w(t)\|^2_c} \left [e^{cx} w(x, t) - \int_{x}^{1} e^{cy} \hat{k}(y-x, t)w(y, t) dy \right ] w(0, t), 
% \end{eqnarray}
but with the exact backstepping transformation \eqref{eq:w_def} and the exact kernel \eqref{eq-exact-adapt-Volterra-eqn} replaced, respectively, by the approximate transformation \eqref{eq-bkst-w-khat} and the DeepONet kernel $\hat{\mathcal{K}}\left(\hat\beta\right)$. 

We use parameter projection for two reasons. One is for ensuring global stability as in exact adaptive PDE control \cite{Smyshlyaev2010}. The second reason, novel in this paper, is for ensuring that the condition of Theorem \ref{theorem:NO_k} remains valid, namely, that  $\|\hat{\beta}\|_{\infty} \leq B$ holds for all time. 
{%\blue 
The Lipschitzness of $\hat\beta(\cdot,t)$, a technical condition for the Arzela-Ascoli theorem and the compactness of the input set of $\mathcal{K}$, seems impossible to enforce without sacrificing the other more important properties enforced by projection, so we assume it instead.
}
%{%\red 
%A slight modification is also made in the projection operator \eqref{eq:weird_proj_def} in order to ensure that $\hat{\beta}$ remains uniformly Lipschitz, which is achieved by keeping $\|\hat{\beta}_x(\cdot, t)\|_{\infty} \leq B_L$ with projection for all $t \geq 0$.}

The elementary pointwise-in-$x$ projection operator \eqref{eq:proj_def} 
%used in \eqref{eq-update-betahat-approx} 
has the following well-known properties \cite[Lemma 8.2]{Smyshlyaev2010},
\begin{align}
\label{eq:tau_square_ppty}
%\label{eq-proj-bound}
 \bullet & \left( \rm{Proj}(\tau, \hat{\beta})\right )^2 \leq \tau^2, \quad \forall (\tau, \hat{\beta}) \in \mathbb{R} \times [0, B] 
  \\
\label{eq-proj-invariance}
 \bullet &  \text{ If } \hat{\beta}(x, 0) \in [-B, B], \forall x \in [0, 1] \text{ then the update law } \nonumber  \\ & \text{  }\hat{\beta} _t \text{ ensures that }\hat{\beta} \in [-B, B]
 \\
\label{eq:tau_proj_inequality}
%\label{eq-proj-inner-prod} 
\bullet &- \tilde{\beta} {\rm Proj}(\tau, \hat{\beta}) \leq - \tilde{\beta} \tau \mbox{ for all } \hat{\beta}, \beta \in [-B, B]\,. 
\end{align}
% \begin{itemize}
%   \item $\left( \rm{Proj}(\tau, \hat{\beta})\right )^2 \leq \tau^2, \quad \forall (\tau, \hat{\beta}) \in \mathbb{R} \times [0, B] $
%   \item If $\hat{\beta}(x, 0) \in [0, B], \forall x \in [0, 1]$ then the update law $\hat{\beta} _t$ ensure that $\hat{\beta}$ remains in $[0, B]$
%   \item $- \tilde{\beta} {\rm Proj}(\tau, \hat{\beta}) \leq - \tilde{\beta} \tau$, for all $\hat{\beta}, \beta \in [0, B]$\,.
% \end{itemize}
%see \cite{Smyshlyaev2010}.
{%\red 
% For the projector \eqref{eq:weird_proj_def}, the properties \eqref{eq:tau_square_ppty}, \eqref{eq-proj-invariance} hold, whereas the property \eqref{eq:tau_proj_inequality} is assumed.\footnote{\blue Property \eqref{eq:tau_proj_inequality} may be violated with $\hat\beta \tau>0$ when $\hat\beta_x =\pm B_L$ and $\hat\beta \tau_x >0$ for some $(x,t)$. In plain terms, the desire to keep $\hat\beta$ Lipschitz in $x$, for the sake of maintaining the approximation accuracy of $\hat{\mathcal{K}}(\hat\beta)$, may result in the update law failing to update $\hat\beta$ when it should update it for the purpose of maintaining stability.}
%This last point is a major assumption, since it assumes that all the classical parameter projection properties are still valid (\eqref{eq:tau_square_ppty} and \eqref{eq:tau_proj_inequality}) .
}

\section{Lyapunov Analysis}\label{section:lyap_computations}

In this section we prove Theorem \ref{thm-main}. 
% The idea is now to replace the {\em exact adaptive} feedback law \eqref{eq:U_def_true_k} with an approximated kernel $\hat{k}$ for an accuracy $\epsilon > 0$ provided by Theorem \ref{theorem:NO_k}, obtaining the {\em approximate adaptive} feedback law
% \begin{equation}
%   U(t)  = \int_{0}^{1} \hat{k}(1-y, t)u(y, t)dy = \hat{k} * u (1, t)= \hat{\mathcal{K}}(\hat\beta)*u(1,t).
%   \label{eq:U_def}
% \end{equation}
We replace backstepping transformation defined in \eqref{eq:w_def} with its approximate version
\begin{equation}
  w(x, t) = u(x, t) - \hat{k} * u (x, t), \quad \forall (x, t) \in [0, 1] \times \mathbb{R}^+\,,\label{eq:wt_def_lyap}
\end{equation}
%Replacing the feedback law \eqref{eq:U_def} with the new one  $U(t) = \int_{0}^{1} \hat{k}(1-y, t)u(y, t)dy$ gives the following 
obtaining (see Appendix \ref{section:appendix_beta_computations}) the pertured target system
\begin{eqnarray}
  w_t(w, t) &=& w_x(x, t) +  \left [\tilde{\beta}(x, t) - \hat{k} * \tilde{\beta} (x, t) \right ] w(0, t) \nonumber \\ && - \Omega(x, t) + w(0, t) \delta(x, t)\,, \label{eq:wt_lyap} \\
  w(1, t) &=& 0. 
  %, \quad \forall (x, t) \in [0, 1] \times \mathbb{R}^+. 
  \label{eq:w1_lyap}
\end{eqnarray}
where 
\begin{eqnarray}
  \tilde{k} &:=& \breve{k} - \hat{k}\,, \\
  \tilde{\beta} &=& \beta - \hat{\beta}\,, \\
  \hat{l} &=& -\hat{k} + \hat{k} * \hat{l} = \mathcal{K}\left(\hat k\right)
  = \mathcal{K}\circ \hat{\mathcal{K}}\left(\hat\beta\right)\,, \\
  \delta &:=& -\tilde{k} + \hat{\beta} * \tilde{k}\,, \\
  \Omega &=&  \hat{k}_t * (w - \hat{l} * w)\,.
\end{eqnarray}
% The computations are done in Appendix \ref{section:appendix_beta_computations}

Before commencing our Lyapunov computations, we  introduce a  lemma on the inverse backstepping kernel $\hat l$.

\begin{lemmma}
\label{lemma_inverse_kernel_properties}
{\em [Inverse kernel properties.]}
Let $B > 0, \hat{\beta} \in \mathcal{C}^0([0, 1] \times \mathbb{R}^+, \mathbb{R})$ such that $\|\hat{\beta}\|_{\infty} \leq B$ and consider 
 the Volterra equation
 \begin{equation}
   \hat{l}(x, t) = -\hat{k}(x, t) + \hat{k} * \hat{l} (x, t), \quad (x, t) \in [0, 1] \times \mathbb{R}^+\,, 
 \end{equation}
with the solution $\hat l =  \mathcal{K}\circ \hat{\mathcal{K}}\left(\hat\beta\right)
%=- \mathcal{K}\left(\hat k\right)
$, where $\hat{k}= \hat{\mathcal{K}}\left(\hat\beta\right)$ is defined with $\hat{\mathcal{K}}$ provided by Theorem \ref{theorem:NO_k} for  accuracy $\epsilon > 0$.  Then %it satisfies the following
 \begin{eqnarray}
   \|\hat{l}\|_{\infty} &\leq& \bar{k} e^{\bar{k}} \,, \label{eq:l_bound} \\
  \bar{k} &:=&  Be^B + \epsilon\,.
\end{eqnarray}
Furthermore, \eqref{eq:w1_lyap} holds if and only if 
 \begin{eqnarray}
   u(x, t) &=& w(x, t) - \hat{l} * w (x, t) \,, 
   \label{eq:l_ppty}
\end{eqnarray}
for any pair of functions $(u,w)$, and in particular when the state $u$ is governed by \eqref{eq:ut_def}, \eqref{eq:u1_def}, and the transformed state $w$ is defined by \eqref{eq:wt_lyap}, \eqref{eq:w1_lyap}.
\end{lemmma}

 \begin{pf}
   The existence of $\hat{l}$ follows from the facts that it satisfies a Volterra integral equation and that $\hat k$ is continuous. The  bound \eqref{eq:l_bound} is obtained with the succesive approximation method, as in the proof of Lemma \ref{lemma_k_existence_and_bound} using \eqref{eq:kernel_bound}.
   To obtain \eqref{eq:l_ppty}, we invoke Lemma \ref{lem-W}. 
   %compute $w - \hat{l} * w$, and replace $w$ with its definition \eqref{eq:wt_def_lyap}.
 \end{pf}

 \begin{lemmma}
 {\em [Lyapunov estimate for perturbed target system.]}
 \label{lemma:V_bound}
 For all $c, B > 0$, there exist strictly positive quantities $\epsilon_0(c, B) = \mathcal{O}_{c \to \infty}(ce^{-\frac{c}{2}}), \gamma_0 = \mathcal{O}_{c \to \infty}(e^{-c})$ with a decreasing dependence on $B$ such that for any $(\epsilon, \gamma) \in (0, \epsilon_0) \times (0,\gamma_0)$, %for any  conditions,
 any $\beta, \hat{\beta}(\cdot, 0) \in \mathcal{C}^0([0, 1])$ that are Lipschitz and satisfy
 \begin{eqnarray}
   \|\beta\|_{\infty}, \|\hat{\beta}(0, \cdot)\|_{\infty} \leq B\,, 
 \end{eqnarray}
 and for any approximate adaptive backstepping kernel $\hat{k}= \hat{\mathcal{K}}\left(\hat\beta\right)$ provided by Theorem \ref{theorem:NO_k} with accuracy $\epsilon$, the perturbed target system \eqref{eq:wt_lyap}, \eqref{eq:w1_lyap} along with the update law \eqref{eq-update-betahat-approx} satisfies 
 \begin{eqnarray}
 \label{eq-bound-betahat}
   |\hat{\beta}(x, t)| &\leq& B,  \quad (x, t) \in [0, 1] \times \mathbb{R}^+\,,  \\
   \dot{V}(t) &\leq& -\frac{c}{4} \frac{\|w(t)\|_c^2}{1 + \|w(t)\|_c^2} - \frac{1}{8} \frac{w^2(0, t)}{1 + \|w(t)\|_c^2}\,, 
    \label{eq-bound-Vdot}
 \end{eqnarray}
 where
 \begin{eqnarray}
   V(t) &:=& \frac{1}{2}
   %\left[
   \ln\left(1 + \|w(t)\|_c^2\right) + \frac{1}{2\gamma} \int_{0}^{1}  \tilde{\beta}^2 (x, t) dx\,,
   %\right]
    \label{eq:V_def} \\
   \|w(t)\|_c^2 &:=& \int_{0}^{1} e^{cx} w^2(x, t) dx\,,
 \end{eqnarray}
 for $t \geq 0$. 
 \end{lemmma}
 
 \begin{pf}
%  For $c, \gamma > 0$, we use the Lyapunov function $V$ defined in \eqref{eq:V_def}.
The property \eqref{eq-bound-betahat} is immediate, as a result of using projection. With the update law \eqref{eq-update-betahat-approx}, taking the derivative of \eqref{eq:V_def} one arrives at %gives after some computations for $t \geq 0$
\begin{eqnarray}
  \label{eq:vdot_bound}
  \dot{V}(t) &=& \frac{1}{1 + \|w(t)\|_c^2} (I_1(t) + I_2(t) + I_3(t) ) 
\nonumber\\
&&+\int_0^1
\underbrace{\frac{w(0,t)e^{cx}\left(1-\hat k*\right)\tilde\beta(x,t)}{1+\|w(t)\|_c^2}}_{=\tau\tilde\beta} dx
\nonumber\\
&&+\frac{1}{\gamma}\int_0^1
\underbrace{\left(-\dot{\hat\beta}(x,t)
\tilde\beta(x,t)\right) }_{\leq -\tau\tilde \beta \mbox{\scriptsize \ \ from \eqref{eq-update-betahat-approx} and \eqref{eq:proj_def}
%\eqref{eq-proj-inner-prod}
}}dx
\nonumber\\  
&\leq& \frac{1}{1 + \|w(t)\|_c^2} (I_1(t) + I_2(t) + I_3(t) )\,,
\end{eqnarray}
where 
\begin{eqnarray}
  I_1(t) &:=& w(0, t)\int_{0}^{1} e^{cx} w(x, t) \delta(x, t)\,, \label{eq:I_1_def} \\
  I_2(t) &:=& - \int_{0}^{1}e^{cx} w(x, t) \Omega(x, t)dx\,, \label{eq:I_2_def}\\
  I_3(t) &:=& - \frac{1}{2} w^2(0, t) - \frac{c}{2} \|w(t)\|^2_c \,.\label{eq:I_3_def}
\end{eqnarray}
Using Lemmas \ref{lemma_inverse_kernel_properties} and \ref{lemma_k_existence_and_bound}, as well as Theorem \ref{theorem:NO_k}, we have the following upper bounds
\begin{eqnarray}
  \|\tilde{k}\|_{\infty} &\leq& \epsilon \label{eq:k_tilde_inequality}\,, \\
  \|\hat{k}\|_{\infty} &\leq& \epsilon + \|\breve{k}\|_{\infty} \leq \epsilon + Be^B =: \bar{k} \,, \\
  \|l\|_{\infty} &\leq& \bar{k}e^{\bar{k}} =: \bar{l} \,, \\
  \|\delta\|_{\infty} &\leq& \epsilon (1 + B) =: \bar{\delta} \epsilon \label{eq:delta_bound} \,, \\
  \|\hat{k}_t(t)\| &\leq& \|\breve{k}_t(t)\| + \epsilon \leq M \|\hat{\beta}_t(t)\| + \epsilon \label{eq:kt_weak_bound}\,, 
\end{eqnarray}
where
\begin{eqnarray}
  \bar{\delta} &:=& 1 + B\,,  \\
  M &:=& 1 + Be^B(2 + Be^B)\,, 
\end{eqnarray}
as well as 
\begin{eqnarray}
  \|\hat{\beta}_t(t)\| &\leq&  \frac{\gamma}{2} \ \frac{e^{\frac{c}{2}}}{1 + \|w(t)\|_c^2} (w^2(0, t) + \|w(t)\|^2_c)(1 + \bar{k}) \label{eq:beta_t_l2_bound}\,.
\end{eqnarray}

\underline{$I_1(t)$ estimate:}
Using \eqref{eq:delta_bound} as well as Young's and Cauchy-Schwarz inequalities, we have the following
\begin{equation}
  I_1(t) \leq \frac{w^2(0, t)}{4} +\epsilon^2 \bar{\delta}^2  e^c \|w(t)\|^2_c \label{eq:I_1_bound}\,.
\end{equation}

\underline{$I_2(t)$ estimate:}
We first rework the upper bound \eqref{eq:kt_weak_bound} using \eqref{eq:beta_t_l2_bound} 
\begin{eqnarray}
  \|\hat{k}_t(t)\| &\leq& \frac{\gamma}{2} \times \frac{e^{\frac{c}{2}} w^2(0, t)}{1 + \|w(t)\|^2_c} \bar{k}_{t} + \frac{\gamma}{2} \times \frac{ e^{\frac{c}{2}} \|w(t)\|^2_c}{1 + \|w(t)\|_c^2} \bar{k}_{t} + \epsilon, \nonumber \\ && \quad t \geq 0\,, \label{eq:kt_true_bound} \\
  \bar{k}_{t} &:=& M (1 + \bar{k})\,.
\end{eqnarray}
Then we use Cauchy-Schwarz inequality to have the following for $t \geq 0$
\begin{eqnarray}
  \|\Omega(\cdot, t)\|_{\infty} &\leq& \|\breve{k}_t(t)\| . \|w(t)\| (1 + \bar{l}) + \epsilon \|w(t)\| (1 + \bar{l})  \nonumber \\
                 &\leq& \gamma e^{\frac{c}{2}}  \frac{w^2(0, t)}{2} \times \frac{\|w(t)\|_c}{1 + \|w(t)\|_c^2} \bar{\Omega} \nonumber \\\nonumber  && + \gamma e^{\frac{c}{2}} \frac{\|w(t)\|}{2} \times \frac{\|w(t)\|_c^2}{1 + \|w(t)\|_c^2} \bar{\Omega} \\ 
                 &&+ \epsilon \|w(t)\| (1 + \bar{l})\,,  \\
  \bar{\Omega} &=:& \bar{k}_t (1 + \bar{l}).
\end{eqnarray}
With these new inequalities we then have the upper bound for \eqref{eq:I_2_def} using Cauchy-Schwarz inequality, 
\begin{eqnarray}
  I_2(t) &\leq&  \gamma e^c \frac{w^2(0, t)}{2} \bar{\Omega} + \gamma e^c \frac{\|w(t)\|_c^2}{2}\bar{\Omega} \nonumber \\ && + \epsilon e^{\frac{c}{2}} \|w(t)\|_c^2 (1 + \bar{l})\,,  \label{eq:I_2_bound}
\end{eqnarray}
where we have used the fact  that $\frac{\|w(t)\|_c^2}{1 + \|w(t)\|_c^2} \leq 1$. 

Finally, gathering \eqref{eq:I_1_bound}, \eqref{eq:I_2_bound}, \eqref{eq:I_3_def}  we have that
\begin{eqnarray}
  \dot{V}(t) &\leq& -\frac{\|w(t)\|_c^2}{1 + \|w(t)\|_c^2} \nonumber \\ && \times \left ( \frac{c}{2}  - \epsilon^2 \bar{\delta}^2 e^c- \gamma e^c \frac{\bar{\Omega}}{2} - \epsilon e^{\frac{c}{2}} (1 + \bar{l}) \right ) \label{eq:wtc_bound} \\
             && -\frac{w^2(0, t)}{1 + \|w(t)\|_c^2} \left ( \frac{1}{4} -  \frac{\gamma e^c}{2} \bar{\Omega} \right)\,. \label{eq:w2_0}
\end{eqnarray}
Noting that the quantities $\bar{\Omega}, \bar{l}$ depend on $\epsilon$ in an increasing fashion, for setting the upper bound on $\gamma, \epsilon$ we fix $\bar{\Omega} := \bar{\Omega}(\epsilon=1), \, \bar{l} :=\bar{l}(\epsilon=1)$. With such fixed choice of $\bar{\Omega}$ and $\bar l$, all the previous inequalities are valid for all $\epsilon \leq 1$.
We now introduce the quantities
\begin{eqnarray}
  \gamma_1 &:=& \frac{ce^{-c}}{4\bar{\Omega}} > 0\,, \label{eq:gamma_1_def} \\
  \epsilon_0 &:=& {\rm min} \left \{ 1, \alpha^{-1}\left(\frac{c}{8}\right) \right \} > 0 \label{eq:eps_0_def}\,, 
\end{eqnarray}
where we introduced the polynomial function $\alpha(\epsilon) = \epsilon^2 \bar{\delta}^2 e^c + \epsilon e^{\frac{c}{2}} (1 + \bar{l})$.
Thus, if we choose $\gamma \in (0, \gamma_1)$ and $\epsilon \in (0, \epsilon_0)$ we get that \eqref{eq:wtc_bound} is dominated by $\leq - \frac{c}{4} \frac{\|w(t)\|_c^2}{1 + \|w(t)\|_c^2}$.
To finish the proof of the lemma, we now consider \eqref{eq:w2_0} and introduce the quantity
\begin{eqnarray}
  \gamma_0 &:=& {\rm min} \left \{ \gamma_1, \frac{e^{-c}}{4 \bar{\Omega}}  \right \} > 0 \label{eq:gamma_0_def} \,.
\end{eqnarray}
Taking $\epsilon \in (0, \epsilon_0), \gamma \in (0, \gamma_0)$ gives
\begin{eqnarray}
  \dot{V}(t) &\leq& -\frac{c}{4} \frac{\|w(t)\|_c^2}{1 + \|w(t)\|^2_c} - \frac{1}{8} \frac{w^2(0, t)}{1 + \|w(t)\|_c^2} \label{eq:vdot_final_bound}\,.
\end{eqnarray}
which completes the proof of \eqref{eq-bound-Vdot}. 
\end{pf}

We are now ready to conclude the proof of Theorem \ref{thm-main}.

\begin{pf}[Proof of Theorem \ref{thm-main}.]
  Let $V$ be the Lyapunov function defined in \eqref{eq:V_def} and $(\epsilon, \gamma) \in (0, \epsilon_0) \times (0, \gamma_0)$, where $\epsilon_0, \gamma_0$ are defined in the proof of Lemma \ref{lemma:V_bound}. It follows from this lemma that $V(t)$ is bounded by $V(0)<\infty$. 
  %, is non-increasing and positive, it is then bounded. 
  From the definition of $V$ as \eqref{eq:V_def}, we  have that $\|\hat{\beta}\|, \|w\|$ are bounded. And by integrating \eqref{eq:vdot_final_bound} in time and keeping in mind that $V$ is nonnegative, we have the following properties in the sense of norms with respect to time:
\begin{itemize}
  \item $\|w\| \in \mathcal{L}_2 \cap \mathcal{L}_{\infty}$
  \item $w(0, \cdot) \in \mathcal{L}_2$
\end{itemize}
%The current result is not enough t
To achieve the convergence of $w$ to $\tilde{0}$, both pointwise and in $ \mathcal{L}_2$, and without seeking an $H^1$ estimate for $w$, borrowing from the approach in (Anfiinsen-Aanmo chapter 4), we introduce the quantity
\begin{eqnarray}
    \alpha(x, t) = \mathcal{B}(u(x,t), k(x)), \quad (x, t) \in [0, 1] \times \mathbb{R}^+ \label{eq:alpha_def_lyap}\,, 
\end{eqnarray}
where $k := \mathcal{K}(\beta)$ is the exact backstepping kernel. %Using Lemma \ref{lemma_k_existence_and_bound}
It follows from \eqref{eq:alpha_def_lyap} that $\alpha$ is solution to the following transport PDE
\begin{eqnarray}
    \alpha_t(x, t) &=& \alpha_x(x, t) , \qquad\qquad (x, t) \in [0,1) \times \mathbb{R}^+\,, \label{eq:alpha_t_def_lyap} \\
    \alpha(1, t) &=& \int_0^1 (\hat{k}(1-y, t) - k(1-y))u(y, t) dy\,. \label{eq:alpha_1_def_lyap}
\end{eqnarray}
From Lemma \ref{lemma_k_existence_and_bound}, it follows that
\begin{equation}
    \|k\|_{\infty} \leq Be^B.
\end{equation}
Through the method of characteristics it follows that
\begin{eqnarray}
    \alpha(x, t) &=& \alpha(1, t+x-1), \quad 
    x+t\geq 1\,.
%    (x, t) \in [0,1] \times \mathbb{R}^+ 
\end{eqnarray}
{%\blue 
and, for $t + x < 1$, we have $\alpha(x, t) = \alpha_0(t+ x)$, where $\alpha_0$ is bounded and denotes the initial condition: $\alpha_0 := u_0 - k * u_0$}.
We thus have that 
\begin{eqnarray}
  |\alpha(x, t)| \leq (Be^B + \bar{k})(1 + \bar{l})
\| w(t+x-1)\|\,,
% \left(\int_0^1 w^2(t+y-1)dy\right)^{1/2}, 
\quad 
%      x+t\geq 1\,.
t + x \geq 1\,,
\label{eq:alpha_infty_bound_lyap}
\end{eqnarray}
{%\blue We thus have that 
and hence $||\alpha||_{\infty} \in \mathcal{L}_{\infty}$
since we've previously shown that $||w|| \in \mathcal{L}_{\infty}$.}
Since the transformation \eqref{eq:alpha_def_lyap} is invertible, 
\begin{equation}
    u = \alpha - \beta * \alpha\,,\label{eq:u_alpha_def}
\end{equation}
and hence we both have $||u||_{\infty} \in \mathcal{L}_{\infty}$ and
\begin{eqnarray}
  |u(x, t)| &\leq& (1+B)(Be^B + \bar{k})(1 + \bar{l})
\|w(t+x-1)\| \,,
% \left(\int_0^1 w^2(t+y-1)dy\right)^{1/2}, 
\nonumber \\ && 
%      x+t\geq 1\,.
t + x \geq 1\,.
% \|w(t+x-1)\|, \quad 
%       x+t\geq 1\,.
% %(x, t) \in [0, 1] \times \mathbb{R}^+ \,.
\label{eq:u-from-alpha_infty_bound_lyap}
\end{eqnarray}
We now  prove that $\|w\| \underset{t \to \infty}{\to} 0$ in order to ultimately obtain $\|u\| \underset{t \to \infty}{\to} 0$.
%to ensure \eqref{eq:u_infty_conv}. 
Since we already know that $\|w\| \in \mathcal{L}_2$, in order to use Barbalat's lemma, we prove that $\|w\|$ is  is uniformly continuous by proving that $\frac{d}{dt}\|w\|^2$ is bounded. We first derive the bound
\begin{eqnarray}
  \left | \frac{d\|w\|^2}{dt}(t) \right | &\leq&  \frac{w^2(0, t)}{2} + 2B\|w(t)\| + 2\bar{k}B \|w(t)\| |w(0, t)| \nonumber \\ && + \gamma e^{\frac{c}{2}} \frac{w^2(0, t)}{2} \bar{\Omega} \nonumber + \gamma e^{\frac{c}{2}} \frac{\|w(t)\|^2}{2} \bar{\Omega} \nonumber \\ && + \epsilon \|w(t)\|^2 (1 + \bar{l}) + \bar{\delta} \epsilon |w(0, t)| \|w(t)\| \,.
\label{eq:w_l2_bound}
\end{eqnarray}
Then, recalling that $\|w\| \in \mathcal{L}_{\infty} \cap \mathcal{L}_2$ and that $w(0, t)=u(0, t)$ is bounded, 
%from the previous remark 
we have that \eqref{eq:w_l2_bound} is bounded. The convergence of $\| w(t) \|$ to zero as time goes to infinity follows from Barbalat's lemma.
From \eqref{eq:u-from-alpha_infty_bound_lyap}, 
%from \eqref{eq:alpha_infty_bound_lyap} it is enough to show that $\|w(t)\| \underset{t \to \infty}{\to} 0$ to have the uniform convergence
\begin{equation}
  \|u(\cdot, t)\|_{\infty} \underset{t \to \infty}{\to} 0 \label{eq:u_infty_conv}\,.
\end{equation}
We now prove the global stabilty \eqref{eq-globstab} in the norm \eqref{eq:gamma_def}. 
% For the global stability we introduce the quantity
% \begin{eqnarray}
%   \Gamma(t) &:=& \|u(t)\|^2 + \|\tilde{\beta}(t)\|^2, 
% \label{eq:gamma_def}
% \end{eqnarray}
Recalling the Lyapunov functional \eqref{eq:V_def},
%let's notice that the following holds
\begin{eqnarray}
  \|w(t)\|^2 &\leq& (e^{2V(t)} - 1), \label{eq:w_l2_V} \\
  \|\tilde{\beta}(t)\|^2 &\leq& 2 \gamma V(t) \leq \gamma (e^{2 V(t)} - 1), \quad t \geq 0 \label{eq:tilde_beta_l2_V}\,.
\end{eqnarray}
With the inverse backstepping transformation $u = w - \hat{l} * w$ we have the upper bound
\begin{equation}
  \|u(t)\|^2 \leq (1 + \bar{l})^2 \|w(t)\|^2 \,.\label{eq:l2_bound_u}
\end{equation}
Gathering \eqref{eq:l2_bound_u}, \eqref{eq:w_l2_V}, \eqref{eq:tilde_beta_l2_V} we have
\begin{eqnarray}
  \Gamma(t) \leq {\rm max} \left (\gamma, (1 + \bar{l})^2 \right ) \times  (e^{2V(t)} - 1) \,.\label{eq:gamma_bound_V}
\end{eqnarray}
Let's also notice that with the backstepping transformation $w = u - \hat{k} * u$,
\begin{eqnarray}
  \frac{1}{2} \ln(1 + \|w(t)\|^2_c) \leq \frac{1}{2} e^c \|w(t)\|^2 \leq \frac{1}{2} e^c (1 + \bar{k})^2 \|u(t)\|^2,\nonumber \\ 
\end{eqnarray}
which leads to
\begin{eqnarray}
  2V(t) \leq {\rm max} \left (\frac{1}{\gamma}, e^c (1 + \bar{k})^2 \right) \times \Gamma(t) \label{eq:V_bound_gamma}, \quad t \geq 0.
\end{eqnarray}
Gathering \eqref{eq:V_bound_gamma} and \eqref{eq:gamma_bound_V} we have the following
\begin{eqnarray}
  \Gamma(t) &\leq& R (e^{\rho \Gamma(0)} - 1),\quad t \geq 0\,, \\
  R &:=& {\rm max} (\gamma, (1 + \bar{l})^2)\,, \\
  \rho &:=& {\rm max} \left ( \frac{1}{\gamma}, e^c (1 + \bar{k})^2 \right )\,.
\end{eqnarray}
Note that the coefficients $R, \rho$ depend in an  increasing fashion on $\epsilon$. To make them independent of the approximation accuracy $\epsilon$, one can choose $\rho := \rho (\epsilon= 1), R := R(\epsilon = 1)$ and all the results are still valid as long as we train the DeepONet $\hat{\mathcal{K}}$ 
%{\color{red} it should be $\hat{\mathcal{K}}$?} 
for $\epsilon \in (0, {\rm min}(1, \epsilon_0))$.
\end{pf}

%\newpage

\section{A Modular Design with a Passive Identifier}\label{section:modular_design_with_passive_identifier}

In this section we depart from the Lyapunov adaptive design of the previous sections and employ a {\em passive identifier} design instead. For ODEs, this identifier is introduced in \cite[Chapter 5]{krstic1995}. Its first use in adaptive control of PDEs is in \cite[Sections 2.1, 3, and 4]{SMYSHLYAEV20071543}, for parabolic PDEs. The first use of a passive identifier in control of a hyperbolic PDE is in \cite[Chapter 4]{Anfinsen2019Adaptive}.

Compared to the Lypunov design, in which the states of the entire system (the plant and the parameter estimator) are captured in a single Lyapunov function, the passive identifier design neither offers superior performance nor the lowest possible dynamic order. In fact, its dynamic order is increased due to the redundancy of the measured state $u(x,t)$ being estimated by another PDE observer state, $\hat u(x,t)$, whose sole role is in the estimation of the unknown parameter. However, the reward for using this less dynamically efficient approach is that the conditions for the estimation of the gain kernel operator are less stringent and the analysis is freed of the requirement to estimate the time derivative of the approximate kernel.

We start by introducing a passive observer-based identifier. For the $u$-system \eqref{eq:ut_def}, \eqref{eq:u1_def}, linearly parametrized in the functional coefficient $\beta(x)$, as proposed in \cite[(4.5)]{Anfinsen2019Adaptive}, we introduce the observer
\begin{eqnarray}
  \hat{u}_t(x, t) &=& \hat{u}_x(x, t) +\hat{\beta}(x, t) u(0, t) +   \gamma_0 (u(x, t) - \hat{u}(x, t))u^2(0, t) \nonumber \\ &&\label{eq:hat_ut_def}  \\
  \label{eq:hat_u_1_def}
  \hat{u}(1, t) &=& U(t) \,.
  %,  \quad (x, t) \in [0, 1] \times \mathbb{R}^+\,,
\end{eqnarray}
where $\gamma_0>$ and the term $\gamma_0 (u(x, t) - \hat{u}(x, t))u^2(0, t)$ represents a form of nonlinear damping in the observer, which plays the same role as update law normalization (namely, to bound the parameter update rate, $\hat\beta_t(x,t)$), and which was introduced in the $x$-passive scheme in \cite[Section 5.6]{SMYSHLYAEV20071543}. 

% In both the plant \eqref{eq:ut_def}, \eqref{eq:u1_def} and in the observer \eqref{eq:hat_ut_def}, \eqref{eq:hat_u_1_def}, the same control input $U(t)$ appears. We take this input as the observer-base feedback law 
% \begin{equation}
%   U(t) = \int_{0}^{1} \hat{k}(1-y, t) \hat{u}(y, t)dy, \quad t \geq 0\,,
%   \label{eq:identifier-based_U}
% \end{equation}
% where $\hat{k}$ is an approximation of the backstepping kernel $\breve k$ defined in \eqref{eq-kbreve-passive}. 
% For any desired approximation accuracy $\epsilon > 0$ Theorem \ref{theorem:NO_k_identifier} guarantees
% \begin{equation}
%   \|(\hat{k} - \breve{k})(\cdot, t)\|_{\infty} \leq \epsilon, \qquad\forall t \geq 0.
% \end{equation}

For the parameter update law, we employ a slight modification of \cite[(4.6)]{Anfinsen2019Adaptive},
\begin{eqnarray}
  \tau(x, t) &:=& \gamma(u(x, t) - \hat{u}(x, t))u(0, t)\,,  \\
  \hat{\beta}_t(x, t) &:=&  {\rm Proj} \left (\tau(x, t), 
\tau_x(x, t) , 
%\hat{\beta}(x, t), 
%\hat{\beta}_x(x, t) 
\right ), \label{eq:update_law_identifier_based} \\
                      && \nonumber
                      \quad (x, t) \in [0, 1] \times \mathbb{R}^+\,,
\end{eqnarray}
where $\gamma > 0$ and the ${\rm Proj}$ is defined in \eqref{eq:proj_def}.
%\eqref{eq:weird_proj_def}. 
%{\red Exactly like in the Lyapunov approach we still require to have $\hat{\beta}_x$ uniformly bounded, for this reason we reuse the same projector with the same assumptions.}
From \cite[Lemma 4.1]{Anfinsen2019Adaptive}, we get the following result, in which the norms $\mathcal{L}_2$ and $\mathcal{L}_\infty$ are with respect to $t\in[0,\infty)$.

\bigskip\begin{lemmma}\label{lemma:ppties_e}
  {\em [Properties of passive identifier \cite[Lemma 4.1]{Anfinsen2019Adaptive}.]}
  The identifier \eqref{eq:hat_ut_def}-\eqref{eq:hat_u_1_def}, with an arbitrary initial condition $\hat u_0 = \hat u(\cdot,0)$ such that $\|\hat u_0\|<\infty$, along with the update law \eqref{eq:update_law_identifier_based} with an 
  arbitrary Lipschitz initial condition $\hat\beta_0 = \hat{\beta}(\cdot, 0) 
  %\in \mathcal{C}^1([0, 1], \mathbb{R})
  $ such that $\|\hat\beta_0\|_{\infty} \leq B
  %, \|\hat\beta_{0x}\|_{\infty} \leq B_L
  $, 
  guarantees that all solutions
%   , for which 
%   %$\hat\beta_t(\cdot,t)$ remains in $H$ and 
% $\hat\beta(\cdot,t)$ remains Lipschitz
% %\allowbreak $- \tilde\beta \, {\rm Proj} (\tau, \tau_x, \hat{\beta}, \hat{\beta}_x)(\cd, k ot, t)  \leq - \tilde\beta \, \tau(\cdot,t)$ 
% for all time, 
satisfy
  \begin{eqnarray}
    \|\hat{\beta}(\cdot, t)\| &\leq& B, \quad t \geq 0\,,\\
    \|e\| &\in& \mathcal{L}_{\infty} \cap \mathcal{L}_2 \,,\\
    |e(0, \cdot)|, \|e\| |u(0, \cdot)|, \|\hat{\beta}_t\| &\in& \mathcal{L}_2\,, 
  \end{eqnarray}
  where
  \begin{equation}
    e := u - \hat{u}\,.
    \label{eq:e_identifier_def}
  \end{equation}
  \end{lemmma}

Next, we introduce our adaptive control law with a DeepONet-approximated gain. Let us first recall the definition of the exact estimated kernel $\breve{k}= \mathcal{K}(\hat\beta)$ through the solution of the Volterra equation
\begin{eqnarray}\label{eq-kbreve-passive}
  \breve{k} (x, t) &=& -\hat\beta(x, t) + \int_{0}^{x} \hat{\beta}(y, t) \breve{k}(x-y, t)dy, \nonumber \\ &&   \quad (x, t) \in [0, 1] \times \mathbb{R}^+.
\end{eqnarray}
A weaker version of an approximating operator $\hat{\mathcal{K}}$ for the approximate estimator kernel $\hat k = \hat{\mathcal{K}}(\hat\beta)$ suffices as compared to the approximation in Theorem \ref{theorem:NO_k}.

\begin{thm}\label{theorem:NO_k_identifier}
{\em [Existence of a NO to approx. the kernel.]}
For all $\epsilon > 0$ there exists a neural operator $\hat{\mathcal{K}}$ such that for all $\hat{\beta}(\cdot, t) \in H$, for all $\forall x \in [0, 1] $, 
\begin{align}
  &\left|\mathcal{K}\left(\hat{\beta}(\cdot, t)\right)(x) - \hat{\mathcal{K}}\left(\hat{\beta}(\cdot, t)\right)(x)\right| < \epsilon \,.
\end{align}
  \end{thm}

  We are now ready to state an equivalent of Theorem \ref{thm-main}
\bigskip
\begin{thm}{\em [Stability of approximate passive-identifier adaptive backstepping control.]} 
  For all $B, \gamma, \gamma_0 > 0$ there exist $\epsilon_0 := \frac{e^{-\frac{3}{2}}}{\sqrt{2}(1+B)} > 0$ such that for all neural operator approximations $\hat{k}$ of accuracy $\epsilon \in (0, \epsilon_0)$ provided by Theorem \ref{theorem:NO_k_identifier} , 
  % and all $\beta, \hat{\beta}_0 \in \mathcal{C}^1([0, 1], \mathbb{R})$ such that
  % \begin{eqnarray}
  %   \|\beta\|_{\infty}, \|\hat{\beta}_0\|_{\infty} &\leq & B \\
  %   \|\beta_x\|_{\infty}, \|\hat{\beta}_{0x}\|_{\infty} &\leq & B_L \,,
  % \end{eqnarray}
  the plant  \eqref{eq:ut_def},\eqref{eq:u1_def}, in feedback with the adaptive control law
  \begin{equation}
    U(t) = \int_{0}^{1} \hat{k}(x-y, t) \hat{u}(y, t)dy\,,
  \end{equation}
  along with the update law for $\hat{\beta}$ given by \eqref{eq:update_law_identifier_based} 
  with any Lipschitz initial condition $\hat\beta_0 = \hat{\beta}(\cdot, 0) 
  %\in \mathcal{C}^1([0, 1], \mathbb{R})
  $ such that $\|\hat\beta_0\|_{\infty} \leq B
  %, \|\hat\beta_{0x}\|_{\infty} \leq B_L
  $, 
  and the passive observer $\hat{u}$ given by \eqref{eq:hat_ut_def}, \eqref{eq:hat_u_1_def}
  with any initial condition $\hat u_0 = \hat u(\cdot,0)$ such that $\|\hat u_0\|<\infty$, satisfies the following properties for all solutions for which
  %such that $\hat\beta_t(\cdot,t)$ remains in $H$
  %, $\hat k(\cdot,t)$ is differentiable, 
%and \allowbreak 
%$- \tilde\beta \, {\rm Proj} (\tau, \tau_x, \hat{\beta}, \hat{\beta}_x)(\cdot, t)  \leq - \tilde\beta \, \tau(\cdot,t)$ 
$\hat\beta(\cdot,t)$ remains Lipschitz for all time:
  \begin{align}
    &\|u\|, \|\hat{u}\|, \|u\|_{\infty}, \|\hat{u}\|_{\infty} \in \mathcal{L}_2 \cap \mathcal{L}_{\infty} \,, \label{eq:theorem_identifier_L} \\
    &\|u\|_{\infty}, \|\hat{u}\|_{\infty} \mapsto 0.\label{eq:theorem_identifier_convergence}  \end{align}
    Additionally, the following global stability estimate holds for the equilibrium $(u,\hat u, \hat\beta) = (0, 0, \beta)$, 
    \begin{equation}
      S(t) \leq R S(0) e^{\rho S(0)}, \quad t \geq 0, \label{eq:s_gs_ppty}
    \end{equation}
    where
    \begin{eqnarray}
      S &:=& \|u\|^2 + \|\hat{u}\|^2 + \|\tilde{\beta}\|^2\,, 
    \end{eqnarray}
    and $\rho, R > 0$ are strictly postive constants.
\end{thm}
\begin{pf}
  The  proof borrows considerably from \cite[Chapter 4]{Anfinsen2019Adaptive}, with minimum repetition, and with necessary augmentation to account for the gain approximation error $\breve k - \hat k$.

  \underline{Part A: Perturbed target system.} We take the same {\em exact} adaptive backstepping transformation as \eqref{eq:w_def} but apply it to the observer state $\hat u$, namely,
\begin{eqnarray}
  w(x, t) &:=& \hat{u}(x, t) - \int_{0}^{x} \breve{k}(x-y, t)\hat{u}(y, t)dy,\nonumber \\ && \quad (x, t) \in [0, 1] \times \mathbb{R}^+ \,, \label{eq:w_identifier_def} 
\end{eqnarray}
where $\breve{k}$ is the exact solution to the Volterra equation \eqref{eq-exact-adapt-Volterra-eqn}. This backstepping transformation leads to the following system satisfied by $w$ (for the computations refer to Appendix \ref{subsection:target_system_identifier_computations}):
\begin{eqnarray}
  w_t(x, t) &=& w_x(x, t) - \breve{k}(x, t) e(0, t) + \nonumber \\ && \gamma_0 u^2(0, t) {\mathcal B}(e,\breve k) (x, t) + \Omega(x, t) \,,\label{eq:wt_identifier_def} \\
  w(1, t) &=& - \int_{0}^{1} \tilde{k}(1-y, t){\mathcal B}(w, \hat\beta)(y, t)dy := \Gamma(t), \nonumber \\ &&  (x, t) \in [0, 1] \times \mathbb{R}^+ \,, \label{eq:gamma_pbc_def}
\end{eqnarray}
where
\begin{eqnarray}
  {\mathcal B}(e,\breve k) &:=& e - \breve{k} * e \label{eq:T_def}\,, \\
  {\mathcal B}(w,\hat\beta) &:=& w - \hat{\beta} * w \label{eq:T_inv_def}\,, \\
  \Omega(x, t) &:=& \int_{0}^{x} \breve{k}_t(x-y, t)  {\mathcal B}(w,\hat\beta)(y, t)dy \,,\\
  \tilde{k} &:=& \breve{k} - \hat{k}\,.
\end{eqnarray}
Notice that the only difference with the system described in \cite[(4.29)]{Anfinsen2019Adaptive} lies in the presence of perturbed boundary conditions $\Gamma$, which is a consequence of the controller choice $U$ that employs an approximated estimated kernel $\hat{k}$ instead of the exact estimated kernel $\breve{k}$.

\underline{Spatial $\mathcal{L}^2$ boundedness and regulation of plant and} \newline \underline{observer states.} We use the following Lyapunov function candidate \cite[(4.42)]{Anfinsen2019Adaptive}:
\begin{equation}
  V(t) := \|w(t)\|_c^2 = \int_{0}^{1}e^{cx}w^2(x, t)dx, \quad t \geq 0\,,
  \label{eq:V_identifier_def}
\end{equation}
where $c>0$ is an arbitrary positive constant. Before starting the Lyapunov computations we first state and recall inequalities that can be achieved from Lemma \ref{lemma_k_existence_and_bound}
\begin{eqnarray}
  \|\breve{k}\|_{\infty} &\leq& Be^B := \bar{k} \label{eq:bar_k_identifier}\,,\\
  \|\tilde{k}\|_{\infty} &\leq& \epsilon \label{eq:tilde_k_identifier}  \,,\\
  \|\breve{k}_t(t)\| &\leq&  M \|\hat{\beta}_t(t)\| \,,\label{eq:kt_identifier_l2_bound} \\
  |\Gamma(t)| &\leq& \epsilon \bar{\Gamma} \|w(t)\| \label{eq:gamma_bound}\,, \\
  \|w(t)\| &\leq& G_1 \|\hat{u}(t)\| \label{eq:w_hatu_bound} \,,\\
  \|\hat{u}(t)\| &\leq& G_2 \|w(t)\| \label{eq:hatu_w_bound}\,,
\end{eqnarray}
where
\begin{eqnarray}
  M &:=& 1 + Be^B(2 + Be^B)\,, \\
  \bar{\Gamma} &:=& 1 + B \,,\\
  G_1 &:=& 1 + \bar{k} \,,\\
  G_2 &:=& 1 + B\,.
\end{eqnarray} 
We use the same computations as the one done in \cite[Chapter 4]{Anfinsen2019Adaptive} with the only difference that $w^2(1, t) = \Gamma^2(t) \neq 0$. We also choose $c=3$ and it leads to the following upper bound
\begin{equation}
  \dot{V}(t) \leq -V(t) \left(1 - e^c \epsilon^2 \bar{\Gamma}^2 \right) + l_1(l) V(t) + l_2(t), \quad t \geq 0\,, \label{eq:V_dot_final_bound_identifier}
\end{equation}
where 
\begin{eqnarray}
  l_1(t) &:=& 2 G_1^2 \gamma_0^2 e^{2c} \|e(t)\|^2 u^2(0, t) + e^c G_2^2 \|\breve{k}_t\|^2 \label{eq:l_1_def}\,, \\
  l_2(t) &:=& (e^3 \bar{k}^2 + 1)e^2(0, t), \quad t \geq 0 \label{eq:l_2_def}\,.
\end{eqnarray}
We introduce
\begin{equation}
  \epsilon_0 :=  \frac{e^{-\frac{c}{2}}}{\sqrt{2}\bar{\Gamma}}\,.
  \label{eq:epsilon_0_identifier}
\end{equation} 
Thus, if we choose $\epsilon \in (0, \epsilon_0)$ we  have $1 - e^c \epsilon^2 \bar{\Gamma}^2  > \frac{1}{2} > 0$.
Since, from \eqref{eq:kt_identifier_l2_bound}, $\|\breve{k}_t\| \leq M \|\hat{\beta}_t\|$, we have from Lemma \ref{lemma:ppties_e} that $l_1, l_2 \in \mathcal{L}^1$ (and are positive). Then using \cite[Lemma B.6]{krstic1995} we have that
\begin{eqnarray}
  V \in \mathcal{L}_1 \cap \mathcal{L}_{\infty}, \quad V(t) \underset{t \to \infty}{\to} 0. \label{eq:V_ppties_identifier}
\end{eqnarray}
It follows from  \eqref{eq:V_ppties_identifier} that $\|w\| \in \mathcal{L}_2 \cap \mathcal{L}_{\infty},  \|w(t)\| \underset{t \to \infty}{\to} 0$. And also from \eqref{eq:hatu_w_bound} we have the same for $\hat{u}$. From Lemma \ref{lemma:ppties_e} it follows that $\|u\| \in \mathcal{L}_2 \cap \mathcal{L}_{\infty}$.

\underline{Part B: Pointwise-in-space boundedness and regulation.} Exactly like in \cite[(3.11)]{Anfinsen2019Adaptive} we also introduce the quantity 
\begin{eqnarray}
  \alpha(x, t) &=& u(x, t) - \int_{0}^{x} k(x-y)u(y, t)dy\,, \nonumber \\ &&   \quad (x, t) \in [0, 1] \times \mathbb{R}^+ \label{eq:alpha_def}
\end{eqnarray}
with $k$ being the exact backstepping kernel i.e $k = \mathcal{K}(\beta)$
The backstepping transformation of \eqref{eq:alpha_def} leads to the following transport PDE 
\begin{eqnarray}
  \alpha_t &=& \alpha_x\,,\label{eq:alpha_t_def} \\
  \alpha(1, t)  &=& \int_{0}^{1}\hat{k}(1-y, t)\hat{u}(y, t)dy - \int_{0}^{1} k(1-y)u(y, t)dy \label{eq:alpha_1_def},\nonumber \\  &&\quad (x, t) \in [0, 1] \times \mathbb{R}^+ \,.
\end{eqnarray}
The only difference with \cite[(4.57b)]{Anfinsen2019Adaptive} lies in the presence $\hat{k}$ instead of $\breve{k}$ in the boundary condition \eqref{eq:alpha_1_def}. But noticing that thanks to \eqref{eq:tilde_k_identifier}, \eqref{eq:bar_k_identifier} we have
\begin{equation}
  |\hat{k}(x, t)| \leq \epsilon + \bar{k}\,,
\end{equation}
and thus $\alpha(1, t)$ remains bounded.  The solution of the transport PDE \eqref{eq:alpha_t_def}-\eqref{eq:alpha_1_def} is given by
\begin{equation}
  \alpha(x, t) = \alpha(1, t +x-1),  \quad x+t\geq 1\,.
  %(x, t) \in [0, 1] \times \mathbb{R}^+\,,
  \label{eq:alpha_value}
\end{equation}
{%\blue 
and, for $t + x < 1$, we have $\alpha(x, t) = \alpha_0(t+ x)$, where $\alpha_0$ is bounded and denotes the initial condition: $\alpha_0 := u_0 - k * u_0$}.
It follows that $\|\alpha\|_{\infty} \in \mathcal{L}_{\infty}$. Since the transformation \eqref{eq:alpha_def} is invertible, $u = \alpha - \beta * \alpha$, we also have that $\|u\|_{\infty} \in \mathcal{L}_{\infty}$. We then achieve an upper bound on $\frac{d}{dt}\|u\|^2$ to get the regulation to $0$ of $\|u\|$ through Barbalat's lemma. From 
\begin{eqnarray}
   \left|\frac{d\|u\|^2}{dt}(t)\right| &\leq& U^2(t) + u^2(0, t) + 2B |u(0, t)| \|u(t)\| < \infty\,, 
\end{eqnarray}
we  have that $\|u(t)\| \underset{t \to \infty}{\to} 0$.
Since $\|u(t)\|, \|\hat u(t)\|\rightarrow 0$, then $\alpha(1,t)\rightarrow 0$.
From the last  observation it follows that 
\begin{eqnarray}
  \|\alpha(\cdot, t)\|_{\infty} \underset{t \to \infty}{\to} 0, \quad t \mapsto \|\alpha(\cdot, t)\|_{\infty} \in \mathcal{L}_2 \cap \mathcal{L}_{\infty}\,.
\end{eqnarray}
With the invertibility of the transformation \eqref{eq:alpha_def}, namely, $u = \alpha  - \beta*\alpha $, we have that
\begin{eqnarray}
  \|u\|_{\infty} \in \mathcal{L}_{\infty} \cap \mathcal{L}_2, \qquad \|u(t)\|_{\infty} \underset{t \to \infty}{\to} 0.
\end{eqnarray}
We now prove a similar result for $\hat{u}$. To do so we first use the change of variable
\begin{equation}
  \hat{e}(x, t) := e(1-x, t)\label{eq:hat_e_def}.
\end{equation}
This leads to the following PDE satisfied  by $\hat{e}$
\begin{eqnarray}
  \hat{e}_t(x, t) + \hat{e}_x(x, t) &=& a(t) \hat{e}(x, t) + f(x, t), \\
  \hat{e}(0, t) &=& 0\,,
\end{eqnarray}
where
\begin{eqnarray}
  f(x, t) &=& \tilde{\beta}(1-x, t) u(0, t)\,,   \\
  a(t) &=& - \gamma_0 u^2(0, t).
\end{eqnarray}
We are now ready to use \cite[Theorem 2.3]{KARAFYLLIS2020104594} to achieve the following ISS result for $\hat{e}$ for $t \geq 1$
\begin{eqnarray}
  \|\hat{e}(\cdot, t)\|_{\infty} &\leq& 2B e^{\left ( 1 + \mu - \gamma_0 \underset{0 \leq s \leq t} {\rm min} u^2(0, s) \right )} \ \underset{t-1 \leq s \leq t} {\rm max}(|u(0, t)|e^{-\mu (t- s)}) \nonumber \\
                          & \leq & 2B e^{1 + \mu} \underset{t-1 \leq s \leq t}{\rm max} |u(0, t)| \label{eq:e_infty_bound}\,, 
\end{eqnarray}
where
\begin{eqnarray}
    \mu &:=& 2 \gamma_0 \ \underset{t \geq 0}{\rm max} \, u^2(0, t) < \infty\,, 
\end{eqnarray}
since $\|u\|_{\infty} \in \mathcal{L}_{\infty}$. From \eqref{eq:e_infty_bound} we are now ready to prove that $\|\hat{e}\|_{\infty} \in \mathcal{L}_2 \cap \mathcal{L}_{\infty}, \, \|\hat{e}\|_{\infty} \underset{t \to \infty}{\to} 0$.
Notice that from \eqref{eq:alpha_def} we have that $u(0, s) = \alpha(0, s)$. And from \eqref{eq:alpha_value} we thus have for $t \leq s \leq t+1$
\begin{eqnarray}
  |u(0, s)| &=& |\alpha(0, s)| = |\alpha(1, s - 1)| = |\alpha(s-t, t)| \nonumber\,,  \\ && \\
  \underset{t \leq s \leq t+1}{\rm max} |u(0, s)| &\leq& \|\alpha(\cdot, t)\|_{\infty} \label{eq:u0_alpha_bound}\,.
\end{eqnarray}
Since $\|\alpha\|_{\infty} \in \mathcal{L}_2, \|\alpha(t)\|_{\infty} \underset{t \to \infty}{\to} 0$, from \eqref{eq:e_infty_bound}, \eqref{eq:u0_alpha_bound} the same holds for $\hat{e}$, and thus for $e$ and the same for $\hat{u}$ since $\hat{u} = u - e$.

\underline{Part C: Global stability.} We now prove \eqref{eq:s_gs_ppty}, that is why we introduce
\begin{equation}
  S(t) := \|u(t)\|^2 + \|\hat{u}(t)\|^2 +\|\tilde{\beta}(t)\|^2, \quad t \geq 0 \label{eq:s_gs_identifier_def}\,. 
\end{equation}
The goal is to prove the existence of a function $\theta \in \mathcal{K}_{\infty}$ such that
\begin{eqnarray}
  S(t) \leq \theta(S(0)), \quad t \geq 0\,.
\end{eqnarray}
For that we reuse the Lyapunov functions introduced in \cite{Anfinsen2019Adaptive} chapter 4 and the reuse the Lyapunov function
\begin{eqnarray}
  V_1(t) &:=& \int_{0}^{1} (1 + x) \left [e^2(x, t) + \frac{1}{\gamma} \tilde{\beta}^2(x, t)dx \right ], \quad t \geq 0 \label{eq:V1_identifier_def}\,.
\end{eqnarray}
The computations made in the proof of  of \cite[Lemma 4.1]{Anfinsen2019Adaptive} leads to the following upper bound : 
\begin{eqnarray}
 \int_{0}^{\infty} e^2(0, \tau) d\tau + \int_{0}^{\infty} \|e(\tau)\|^2d\tau &&\nonumber \\  + 2 \gamma_0 \int_{0}^{\infty} \|e(\tau)\|^2 u^2(0, \tau)d\tau &\leq& V_1(0) \label{eq:e-u0-l2-bound-v1}\,.
\end{eqnarray}
Also from the definition of the udpate law \eqref{eq:update_law_identifier_based} we have that
\begin{equation}
  \|\breve{k}_t(t)\|^2 \leq M^2 \|\hat{\beta}_t(t)\|^2 \leq \frac{M^2 \gamma^2}{2 \gamma_0}  (2 \gamma_0\|e(t)\|^2 u^2(0, t))\,.
  \label{eq:beta_hat_V1_l2_bound}
\end{equation}
Recalling \eqref{eq:V_dot_final_bound_identifier}, we also have from \cite[Lemma B.6]{krstic1995} that
\begin{eqnarray}
  V(t) &\leq& (e^{-\frac{t}{2}}V(0) + \|l_2\|_{1})e^{\|l_1\|_1} \,. \label{eq:V_good_bound_identifier} % I am running out of ideas to name the equations
\end{eqnarray}
Then recalling \eqref{eq:l_1_def}, \eqref{eq:l_2_def} as well as \eqref{eq:beta_hat_V1_l2_bound} and \eqref{eq:e-u0-l2-bound-v1} we have the following
\begin{eqnarray}
  \|l_1\|_1 \leq \bar{l}_1 V_1(0) \label{eq:l_1_bound}\,, \\
  \|l_2\|_2 \leq \bar{l}_2 V_1(0) \label{eq:l_2_bound} \,. 
\end{eqnarray}
where, recalling that $\bar k = B {\rm e}^{B}, G_1 = B {\rm e}^{B}, G_2 = B, M (B)= 1 + Be^B(2 + Be^B)$, the coefficients $\bar l_1, \bar l_2$ are given by
\begin{eqnarray}
\label{eq-l1bar}
  \bar{l}_1(B,\gamma,\gamma_0) &:=& {\rm max} \bigg (
  %G_1
  \big(1+B {\rm e}^B\big)^2 \gamma_0 e^{2c}, \nonumber \\ &&  \frac{\gamma^2 e^c (1 + B)^2 \big( 1 + Be^B(2 + Be^B)\big)^2 }{2 \gamma_0}
  %G_2
   \bigg)\,,   \\
\label{eq-l2bar}
  \bar{l}_2(B) &:=& 1+ e^3 %\bar{k}
  \left(1+B {\rm e}^B\right)^2 .
\end{eqnarray}
%with $M (B)= 1 + Be^B(2 + Be^B)$. 
We then introduce the function
\begin{eqnarray}
  V_3(t) &:=& V_1(t) + V(t) 
   \nonumber \\
&=& \int_{0}^{1}(1+x)\left [\frac{\tilde{\beta}(x, t)^2}{\gamma} + e^2(x, t) \right ]\nonumber \\ && + \int_{0}^{1} e^{3x} w^2(x, t)dx\,.
  \label{eq:V3_def}
\end{eqnarray}
Noticing that
\begin{equation}
  V_1(t) \leq \bar{l}_2 V_1(0) e^{\|l_1\|_1}, \quad t \geq 0\,,
  \label{eq:V1_bound}
\end{equation}
we achieve from \eqref{eq:V_good_bound_identifier}, \eqref{eq:V1_bound}, \eqref{eq:l_1_bound} and \eqref{eq:l_2_bound} the following
\begin{eqnarray}
  V_3(t) \leq 2 \bar{l}_2 V_3(0) e^{\bar{l}_1 V_3(0)}, \quad t \geq 0 \label{eq:V3_bound}.
\end{eqnarray}
We thus have the following for $t \geq 0$ using the Cauchy-Schwarz and Young's inequality
\begin{eqnarray}
  V_3(t) &\geq& \frac{1}{\gamma} \|\tilde{\beta}(t)\|^2 + \|e(t)\|^2 + \frac{\|\hat{u}(t)\|^2}{(1 + B)^2} \nonumber \\
         &\geq& \frac{1}{\gamma}\|\tilde{\beta}(t)\|^2 + \frac{1}{(1 + B)^2} (\|e(t)\|^2 + \|\hat{u}(t)\|^2) \nonumber \\
         &\geq& \frac{1}{\gamma} \|\tilde{\beta}(t)\|^2 + \frac{1}{(1 + B)^2} (\|u(t)\|^2 + 2 \|\hat{u}(t)\|^2 \nonumber \\ && - 2 \|u(t)\|\|\hat{u}(t)\|) \nonumber \\
         &\geq& \frac{1}{\gamma} \|\tilde{\beta}(t)\|^2 + \frac{1}{(1 + B )^2}\left(\frac{1}{4} \|\hat{u}(t)\|^2 + \frac{1}{3} \|u(t)\|^2\right) \nonumber \\
         &\geq& {\rm min}\left ( \frac{1}{\gamma}, \frac{1}{4(1 + B)^2} \right ) S(t) \label{eq:V3_lower_bound_s}.
\end{eqnarray}
We now focus on establishing the upper bound on $V_3$.
From $\eqref{eq:V3_def}$ we have for $t \geq 0$ with Young's inequality
\begin{eqnarray}
  V_3(t) &\leq& \frac{2}{\gamma}\|\tilde{\beta}(t)\|^2 + 4 \|u(t)\|^2 + 4 \|\hat{u}(t)\|^2 + e^{3} (1 + \bar{k})^2 \|\hat{u}(t)\|^2 \nonumber \\
         &\leq &{\rm max} \left (\frac{2}{\gamma}, 4 + e^3 (1 + \bar{k})^2 \right) S(t) \label{eq:V3_upper_bound}\,.
\end{eqnarray}
Then gathering \eqref{eq:V3_bound}, \eqref{eq:V3_lower_bound_s} , \eqref{eq:V3_upper_bound} we obtain \eqref{eq:s_gs_ppty}
with
\begin{eqnarray}
\label{eq-Rbound}
  R (B, \gamma, \gamma_0)&:=& 2 \bar{l}_2 \, {\rm max} \left ( \gamma, 4(1 + B)^2 \right )\,, \\
\label{eq-rhobound}
  \rho(B, \gamma, \gamma_0) &:=& \bar{l}_1 \, {\rm max} \left ( \frac{2}{\gamma}, 4 + e^3\left(1+B {\rm e}^B\right)^2\right )\,.
\end{eqnarray}
\end{pf}

Examining the bounds $R$ and $\rho$ in \eqref{eq-Rbound}, \eqref{eq-rhobound}, 
 in light of \eqref{eq-l1bar} and \eqref{eq-l2bar}, one notes their explicit, albeit conservative dependence on the ``instability bound'' $B$, the adaptation gain $\gamma$, and the normalization (observer nonlinear damping) gain $\gamma_0$. The increasing dependence on the instability $B$ is the most evident, and expected.

\section{Simulations} \label{section:simulations}
\setlength{\parskip}{0.5em}
We simulate the system governed by \eqref{eq:ut_def}, \eqref{eq:u1_def} where the plant coefficient $\beta(x)=5\cos(\sigma \cos^{-1}(x))$ is defined as a Chebyshev polynomial with shape parameter $\sigma$. This choice of $\beta(x)$ follows from \cite{bhan_neural_2023}, \cite{GS-preprint}, but we emphasize that any compact set of continuous functions can be chosen for the plant coefficients $\beta(x)$. For simulation of the hyperbolic PDE, we utilize a first-order upwind scheme with temporal step $dt=5\times 10^{-3}$ and spatial step $dx=1 \times 10^{-2}$. We note that the given PDE with $\beta(x)$ as a Chebyshev polynomial is open-loop unstable (Figure 3, \cite{bhan_neural_2023}). For the adaptive control scheme, we utilize the Lyapunov approach given in \eqref{eq-control}, \eqref{eq-update-betahat-approx}, \eqref{eq-tau-approx}, \eqref{eq-bkst-w-khat} with a first order Euler scheme to simulate \eqref{eq-update-betahat-approx}. 

We now begin our discussion on training the NO-approximated kernel. To effectively handle the adaptive estimates of $\hat{\beta}(x)$ and the corresponding kernels, one must construct a diverse and exhaustive dataset anticipating the possible $\beta$ functions encountered. The simplest way to build this dataset is by generating $\beta$ values with varying $\sigma$ and simulating the true adaptive controller saving both the $\beta$ functions and corresponding kernels encountered. Although simulating the true adaptive controller is expensive, the construction of the dataset, like training, only needs to be done once offline.  As such, for the particular neural operator developed in this work, we considered $10$ $\beta$ functions with $\sigma \sim \text{Uniform}(2.7, 3.2)$ and simulate the resulting PDEs under the adaptive controller for $T=10$s sub-sampling each pair of $(\beta, k)$ every $0.01$s. This creates a total dataset of $10000$ different $(\beta, k)$ pairs to perform supervised learning of the neural operator (available publicly \href{https://drive.google.com/drive/folders/16bztrIFXy700_37LBvdguyTOf8EP0MQy?usp=sharing}{here}). We note that if one wants to handle a larger range of plant coefficients (wider range of $\sigma$ values), they will need to sample more $\beta$ functions and perform similar calculations running the true adaptive controller. Lastly, we briefly mention that although the Lyapunov approach as discussed in Secs. \ref{section:no_ppties}, \ref{sec-Lyap-design} requires approximation of the derivatives, we found sufficient performance without the calculation intensive derivative approximation (More details on neural operator approximation of derivatives can be found in Sec. IX of \cite{GS-preprint}). 

The training of the NO uses the DeepXDE package \cite{lu2021deepxde} and requires approximately $100$ seconds to train (whereas the dataset takes several minutes to construct). The resulting DeepONet consists of $14913$ parameters with traditional multi-layer perceptron (MLPs) for the branch and trunk networks. Despite the small network, excellent accuracy is achieved as the $L_2$ training error was $2 \times 10^{-3}$ and the $L_2$ testing error was $1.8 \times 10^{-3}$. 

We begin our discussion of the numerical simulations by presenting NO speedups averaged over $100$ calculations of the kernel, according to discretization size, in Table \ref{tab:nopspeedups}. We can see that as the spatial step size grows, the speedup gained becomes immense shrinking the analytical kernel calculation time from \textit{4 minutes to 0.4 seconds}. This is only for a single kernel calculation in which the speedup is exemplified at each timestep as in adaptive control, the resulting kernel needs to be continually recalculated according to the new $\hat{\beta}$ estimate. 

\begin{figure*}
    \centering
    \includegraphics{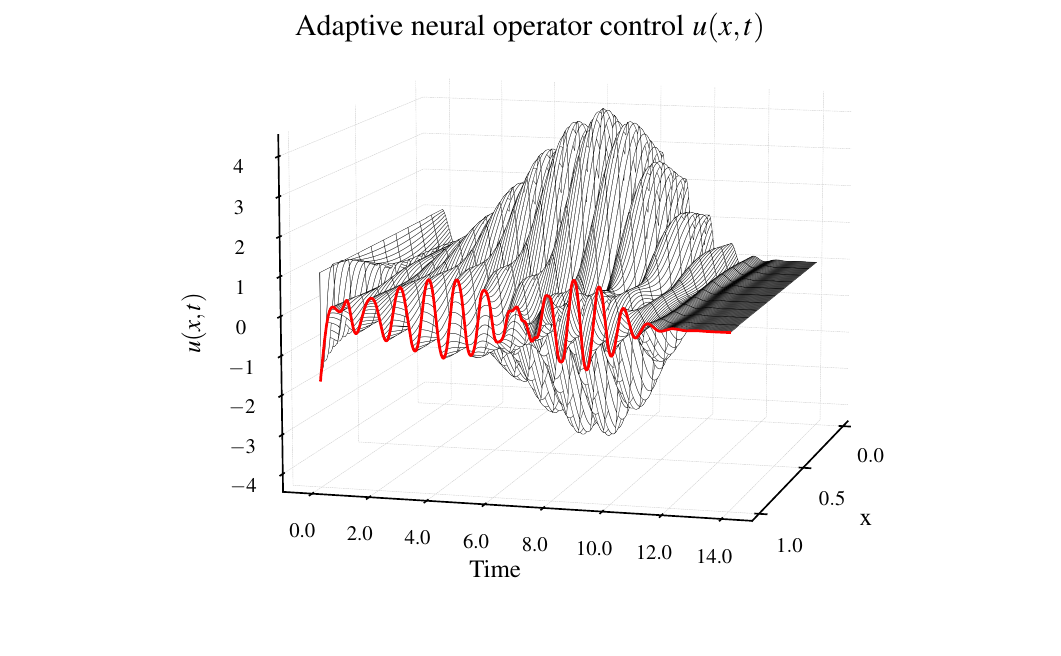}
    \caption{Adaptive neural operator controller applied to the PDE governed by \eqref{eq:ut_def}, \eqref{eq:u1_def} where $\beta(x)=5\cos(\sigma \cos^{-1}(x))$ with $\sigma=2.9$ and initial condition $u(x, 0)=1$. The initial guess for $\hat{\beta}$ was $\hat{\beta}(x, 0)=1 \quad \forall x \in [0, 1]$ and the control update law \eqref{eq-update-betahat-approx}, \eqref{eq-tau-approx}, \eqref{eq-bkst-w-khat}, \eqref{eq-norm-w} has parameter $c=1$.}
    \label{fig:adaptiveControlStabilization}
\end{figure*}

\begin{figure*}
    \centering
    \includegraphics{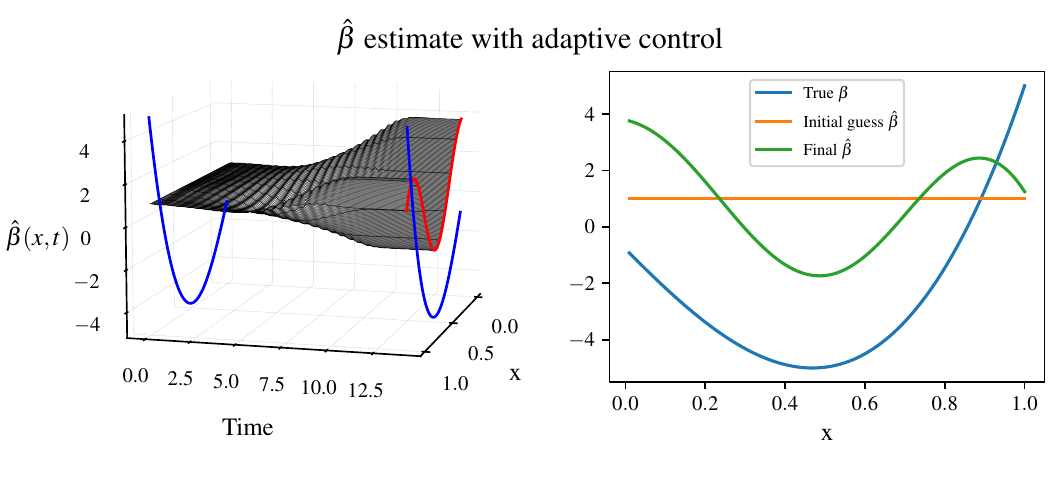}
    \caption{Left: $\hat{\beta}$ estimates when controlling the PDE in Figure \ref{fig:adaptiveControlStabilization} using neural operator approximated kernels: true $\beta$ (blue) and final estimated $\hat{\beta}$ (red); Right: comparison between the true $\beta$ value, the initial guess $\hat{\beta}(\cdot, 0)=1$, and the final estimated $\hat{\beta}$ at $t=13$.}
    \label{fig:betaestimate}
\end{figure*}

\begin{figure*}
    \centering
    \includegraphics{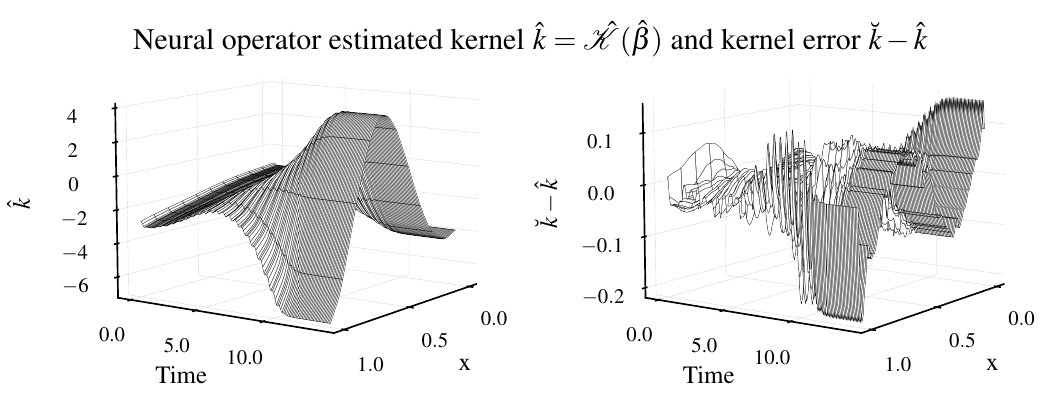}
    \caption{Neural operator approximated kernels when controlling the PDE in Figure \ref{fig:adaptiveControlStabilization}(left), and the difference in kernel error between the approximated kernel and the analytical 
kernel (right).}
    \label{fig:adaptiveControlKernels}
\end{figure*}

Lastly, we conclude by presenting a single instance of  the resulting controller under NO approximated kernels in Figure \ref{fig:adaptiveControlStabilization}. This instance presents $\beta(x)$ as the aforementioned Chebyshev polynomial with $\sigma = 2.9$ and initializes the estimated plant parameter to $\hat{\beta}(x, 0)=1$. We emphasize that this specific $\beta(x)$ was not seen in any of the $\beta(x)$ functions utilized for training. We begin by illustrating the nature of adaptive control explicitly in Figure \ref{fig:adaptiveControlStabilization}. In Figure \ref{fig:adaptiveControlStabilization}, the plant's instability in the first eight seconds drives the estimation of $\hat{\beta}$, but then, by ten seconds, the estimate is good enough to provide a stabilizing controller for the system leading to rapid decay of the system state. Furthermore the stabilization annihilates the persistence of excitation from the plant's estimator leading to the stagnation of the estimate $\hat{\beta}$. This is observed clearly in Figure \ref{fig:betaestimate} where $\hat{\beta}$ freezes by $t=10$ and --- due to lack of excitation --- never reaches the true $\beta(x)$ value. We stress that this lack of convergence towards the true $\beta$ is not an issue but merely a feature of adaptive control as one is not performing perfect plant identification, but estimating with the goal of stabilization, which is aptly achieved with the final, inexact $\hat{\beta}$ (in red) of Figure \ref{fig:betaestimate}. We conclude our discussion with Figure \ref{fig:adaptiveControlKernels} showcasing the kernel computed using the NO over time. As expected, once $\hat{\beta}$ stalls, the corresponding kernel --- which is a mapping relying solely on $\hat{\beta}$ --- stagnates concurrently. Furthermore, in the right of Figure \ref{fig:adaptiveControlKernels} we see that the NO approximation is very close to the analytical estimate maximizing at an error of approximately $10\%$ with respect to the kernels magnitude.

\definecolor{lighter-green}{rgb}{0, 0.6, 0.00392156862} 
\begin{table}[t]
\label{tab:nopspeedups}
\centering
\resizebox{\columnwidth}{!}{%
\begin{tabular}{lccc}
\hline
\textbf{\begin{tabular}[c]{@{}l@{}}Spatial Step \\ Size (dx)\end{tabular}} & \multicolumn{1}{l}{\textbf{\begin{tabular}[c]{@{}l@{}}Analytical \\ Kernel \\ Calculation \\ Time(s) $\downarrow$ \end{tabular}}} & \multicolumn{1}{l}{\textbf{\begin{tabular}[c]{@{}l@{}}Neural Operator\\ Kernel\\ Calculation \\ Time(s) $\downarrow$ \end{tabular}}} & \multicolumn{1}{l}{\textbf{Speedup} $\uparrow$} \\ \hline
$0.01$                                                                     & $0.044$                                                                                                             & $0.023$                                                                                                                & {\color{lighter-green} $1.87$x}       \\
$0.001$                                                                    & $2.697$                                                                                                               & $0.024$                                                                                                                & {\color{lighter-green} $110$x}         \\
$0.0005$                                                                    & $10.334$                                                                                                                & $0.024$                                                                                                                & {\color{lighter-green} $427$x}        \\
$0.0001$                                                                   & $245$                                                                                                               & $0.037$                                                                                                                & {\color{lighter-green} $6642$x}       \\ \hline
\end{tabular}%
} 
\caption{Neural operator speedups over the analytical kernel calculation with respect to the increase in discretization points (decrease in step size). }
\end{table}

\section{Conclusion}
In this paper, we present the first results for NO approximated kernels in adaptive control of hyperbolic PDEs. We consider two approaches, namely a Lyapunov-based approach and a modular approach with a passive identifier, and prove global stability for both approaches, with tradeoffs between assumptions and dynamic orders. We then present numerical simulations showcasing the viability of the Lyapunov approach under the neural operator approximated kernels obtaining speedups on the magnitude of $10^3$. With such large reduction in computational costs, NO-based adaptive backstepping opens the door for applying adaptive PDE control in real-time. 
\setlength{\parskip}{.6em}  

\appendix
\section*{Appendix}
\section{Backstepping Transformation and Involution Operator for the Kernel}
We introduce a {\em backstepping operator} $\mathcal{B}$
%$: (u,k) \mapsto w$ 
defined as
\begin{equation}\label{eq-bkst-oper}
    %w = 
    \mathcal{B}(\xi,\eta) := \xi - \eta *\xi\,, 
\end{equation}
and, with this operator, introduce the (Volterra-type) {\em backstepping equation}
\begin{eqnarray}\label{eq-bkst-xietazeta}
    \mathcal{B}(\xi,\eta) =\zeta \,, 
   % \mathcal{B}(\xi,\eta) + \zeta &=& \nonumber\\
   %  \xi -\eta*\xi + \zeta &=& 0\,. 
\end{eqnarray}
meant to be solved for $\xi$, for given $(\zeta,\eta)$. 
We denote the solution of \eqref{eq-bkst-xietazeta} for $\xi$ with the operator $\mathcal{W}(\zeta,\eta)$.
%namely, as
% \begin{eqnarray}\label{eq-bkst-W}
%     \mathcal{B}(\mathcal{W}(\zeta,\eta),\eta) =
%     \mathcal{W}(\zeta,\eta) - \eta*\mathcal{W}(\zeta,\eta)  = \zeta \,. 
%     % \mathcal{B}(\mathcal{W}(\eta,\zeta),\eta) + \zeta 
%     % &=& \nonumber  \\
%     % \mathcal{W}(\eta,\zeta) - \eta*\mathcal{W}(\eta,\zeta)  + \zeta &=&0\,. 
% \end{eqnarray}
Next, setting $\zeta = -\eta$ in \eqref{eq-bkst-xietazeta}, we introduce the {\em kernel integral equation}
\begin{eqnarray}\label{eq-kernel-eqn}
    \mathcal{B}(\xi,\eta) + \eta= 
    \xi - \eta*\xi +\eta  =  0\,, 
\end{eqnarray}
and denote its solution for $\xi$ with the operator 
%\begin{equation}
    $\mathcal{K}(\eta) :=\mathcal{W}(-\eta,\eta)$,
%\end{equation}
namely, as
\begin{eqnarray}\label{eq-eqn-K}
    \mathcal{B}(\mathcal{K}(\eta),\eta) + \eta &=& \nonumber \\
    \mathcal{K}(\eta) - \eta*\mathcal{K}(\eta)  + \eta &=&0\,. 
\end{eqnarray}

Next, we give a previously unobserved property of $\mathcal K$. 

\begin{lemmma}\label{lem-invol}
$  \mathcal{K}^{-1} = \mathcal{K}$,  namely $ \mathcal{K}^2 := \mathcal{K}\circ \mathcal{K} = \mbox{Id}$.
\end{lemmma}

\begin{pf}
    By noting that the roles of $\xi$ and $\eta$ in \eqref{eq-kernel-eqn} are interchangeable, or by using the Laplace transform. 
\end{pf}

Due to the property given by Lemma \ref{lem-invol}, we call $\mathcal{K}$ the {\em involution operator}.\footnote{Because a matrix $A$ such that $A^2=I$ is typically referred to as {\em involutory}.}

% We highlight another property that trivially follows from the definition of $\mathcal{K}$.

% \begin{lemmma}\label{lem-product}
% $ \eta*\mathcal{K}(\eta) =\mathcal{K}(\eta)  + \eta$.
% \end{lemmma}

The next lemma gives an explicit expression for operator $\mathcal{W}$.
%and an additional property of $\mathcal{W}$. 

\begin{lemmma}\label{lem-W} 
%Operator $\mathcal{W}$ has the following properties:
\begin{eqnarray}
    \mathcal{W}(\zeta,\eta) \nonumber  &=& \zeta - \mathcal{K}(\eta)*\zeta  \\ \nonumber &=& \mathcal{B}\left(\zeta,\mathcal{K}(\eta)\right) \\ 
    &=& \mathcal{B}\left(\zeta,\mathcal{K}^{-1}(\eta)\right)\,.
%    \\    \label{eq-W-explicit}   \eta*\mathcal{W}(\zeta,\eta) &=& -\mathcal{K}(\eta)*\zeta    \,.
\end{eqnarray}
\end{lemmma}

\begin{pf}
    By direct substitution into \eqref{eq-bkst-xietazeta}, or by using the Laplace transform. The last equality 
    %in \eqref{lem-W} 
    follows from Lemma \ref{lem-invol}.
\end{pf}

To summarize,
\begin{equation}
\zeta = \mathcal{B}\left(\xi,\eta\right) \quad \mbox{iff} \quad
\xi=\mathcal{B}\left(\zeta,\mathcal{K}^{-1}(\eta)\right)\,,
\end{equation}
or, alternatively stated, if $\xi + \eta = \eta*\xi$, then
\begin{equation}
w = u-\xi*u \quad \mbox{iff} \quad  u = w-\eta*w\,.
\end{equation}
These observations yield the following result.

\begin{lemmma}
The operator $(\eta,\zeta) \mapsto (\mathcal{K}(\eta),\mathcal{B}(\zeta,\eta))$ is an involution.
\end{lemmma}

\begin{pf} By noting that
\begin{eqnarray}
\mathcal{K}(\mathcal{K}(\eta)) &=& \eta \\
\mathcal{B}\left(\mathcal{B}\left(\zeta,\eta\right),\mathcal{K}(\eta)\right) &=&\zeta\,.
\end{eqnarray}
\end{pf}

In calculations to come, equation \eqref{eq-bkst-xietazeta} will arise in a particular form. We provide its solution in the following lemma. 

\begin{lemmma}\label{lem-k0k1}
For given functions $\beta_0,  \beta_1$, and $k_0=\mathcal{K}(\beta_0)$, if the function $k_1$ satisfies the equation
\begin{equation}\label{eq-k0k1}
k_1 - \beta_0*k_1 +\beta_1 -\beta_1 * k_0 =0\,,
%k_1 - \beta_1*k_1 +\beta_1 -\beta_0 * k_0 =0\,,
\end{equation}
then it is explicitly given by
\begin{eqnarray}\label{eq-k1explicit}
k_1 = \mathcal{K}_1(\beta_0,\beta_1) \nonumber &:=&  -\beta_1 +\beta_1*\mathcal{K}(\beta_0) \\ && + \beta_1*\mathcal{K}(\beta_0)\nonumber \\ && - \beta_1 * \mathcal{K}(\beta_0) * \mathcal{K}(\beta_0)\,.
%k_1 = \mathcal{K}_1(\beta_0,\beta_1) :=  -\beta_1 +\beta_0*\mathcal{K}(\beta_0) + \beta_1*\mathcal{K}(\beta_1) - \beta_0 * \mathcal{K}(\beta_0) * \mathcal{K}(\beta_1)\,.
\end{eqnarray}
\end{lemmma}

\begin{pf}
    Using Lemma \ref{lem-W}. 
\end{pf}
\vspace{-10pt}

\section{Perturbed Target System with Approximate Estimated Kernel}

\label{section:appendix_beta_computations}

We derive the perturbed target system \eqref{eq:wt_lyap}, \eqref{eq:w1_lyap}, where $w =  u - \hat{k} * u$ 
and $\hat{k}$ is the approximate estimated kernel, assumed to be both continuous and {differentiable} with respect to $t$. Since \eqref{eq:w1_lyap} is just a consequence of the choice of $U(t)$, we focus on proving \eqref{eq:wt_lyap}. Taking the derivative with respect to $x$ and $t$ of \eqref{eq:wt_def_lyap} gives the following
\begin{eqnarray}
  w_t &=& u_t - \hat{k}_t * u - \hat{k} * u_t \label{eq:wt_appendix} \\
  w_x(x, t) &=& u_x(x, t) - \hat{k} (0, t) u(x, t) \nonumber \\ && + \int_{0}^{x} \hat{k}_y(x-y, t) u(y, t) dy \nonumber \label{eq:wx_appendix_ongoing}, \\
&&\quad \forall (x, t) \in [0, 1]  \times \mathbb{R}^+. 
\end{eqnarray}
Then doing an integration by parts on \eqref{eq:wx_appendix_ongoing} gives
\begin{equation}
  w_x(x, t)  = u_x(x, t) - \hat{k}(x, t) w(0, t) - \hat{k} * u_x (x, t)
  \label{eq:wx_appendix}\,,
\end{equation}
noticing that $u(0, t) = w(0, t)$. Gathering \eqref{eq:wt_appendix}, \eqref{eq:wx_appendix} and using \eqref{eq:ut_def} gives
\begin{eqnarray}
\label{eq-wtwx}
  w_t(x, t) - w_x(x, t) &=& w(0, t) \bigg [\beta(x) + \hat{k}(x, t) \nonumber \\ && - \int_0^x \hat{k}(x-y, t) \beta (y)dy \bigg ] \nonumber \\
  &&- \hat{k}_t * u(x, t), 
%\\ &&
\quad \forall (x, t) \in [0, 1)  \times \mathbb{R}^+.
\end{eqnarray}
From \eqref{eq-exact-adapt-Volterra-eqn}, we have that 
\begin{eqnarray}
  \hat{k} &=& -\hat{\beta} + \hat{\beta} * \hat{k} + \delta \label{eq:k_almost_def} 
  % \\
  % \delta  &:=& -\tilde{k} + \hat{\beta} * \tilde{k}\\
  % \tilde{k} &:=& \breve{k} - \hat{k} \\
  % \tilde{\beta}(x, t) &:=& \beta(x) - \hat{\beta}(x, t) , \quad \forall (x, t) \in [0, 1] \times \mathbb{R}^+ 
\end{eqnarray}
and, with some rearrangements, arrive at
%From \eqref{eq:k_almost_def} and some computations we have that 
\begin{eqnarray}
  \beta(x) + \hat{k}(x, t) - \int_0^x \hat{k}(x-y, t) \beta(y)dy &=& \tilde{\beta}(x, t) \nonumber \\ && - \tilde{\beta} * \hat{k} (x, t) \nonumber \\ &&  + \delta(x, t)\,.
\end{eqnarray}
Then, using the inverse backstepping transformation $u = w - \hat{l} * w$, from \eqref{eq-wtwx}
% \begin{eqnarray}
%   u &=& w - \hat{l} * w \\
%   \hat{l} &:=& -\hat{k} + \hat{k} * \hat{l}
% \end{eqnarray}
we arrive at \eqref{eq:wt_lyap}. %\eqref{eq:wt_def_lyap}.

\section{Perturbed Observer Target System with Exact Estimated Kernel}
\label{subsection:target_system_identifier_computations}

We derive the system \eqref{eq:wt_identifier_def}-\eqref{eq:gamma_pbc_def}. Since \eqref{eq:gamma_pbc_def} is just a matter of the choice for the controller, we focus on \eqref{eq:wt_identifier_def}. 
% For the reader's convenience we recall the backstepping transformation  \eqref{eq:w_identifier_def} 
% \begin{eqnarray*}
%   w(x, t) = \hat{u}(x, t)  - \int_{0}^{x} \breve{k}(y-x, t)\hat{u}(y, t)dy
% \end{eqnarray*}
Taking the derivative of \eqref{eq:w_identifier_def}  with respect to $t$ gives
\begin{eqnarray}
  w_t(x, t) &=& \hat{u}_t(x, t) - \int_{0}^{x} \breve{k}(x-y, t) \hat{u}_t(y, t)dy - \Omega(x, t) \label{eq:wt_identifier_computations}\,, \\
  \Omega(x, t) &:=& \int_{0}^{x} \breve{k}_t(x-y, t) \hat{u}(y, t)dy\,.
  %,  \quad (x, t) \in [0, 1] \times \mathbb{R}^+
\end{eqnarray}
and with respect to $x$ gives
\begin{eqnarray}
\label{eq:wx_identifier_computations} 
w_x(x, t) &=& \hat{u}_x(x, t) - \breve{k} (0, t) \hat{u}(x, t)  + \int_{0}^{x} \breve{k}_y (y-x, t) \hat{u}(y, t) dy \nonumber \\
            &=& \hat{u}_x(x, t) - \breve{k}(x, t)\hat{u}(0, t) - \int_{0}^{x} \breve{k} (x-y, t) \hat{u}_x(y, t)dy \nonumber \\ && 
%\nonumber\\         && \qquad\qquad \qquad (x, t) \in [0, 1] \times \mathbb{R}^+
\end{eqnarray}
where we used integration by parts. Then gathering \eqref{eq:wt_identifier_computations}, \eqref{eq:wx_identifier_computations} we have
\begin{eqnarray}
  w_t(x, t) - w_x(x, t) &=& \hat{u}(0, t)\breve{k}(x, t) \nonumber \\ && + u(0, t) \left [\hat{\beta}(x, t)  - \int_{0}^{x} \breve{k}(x-y, t) \hat{\beta}(y, t)dy   \right ] \nonumber \\
                        &&- \Omega(x, t)  + \gamma_0 u^2(0, t) \mathcal{B}(e, \breve{k})(x, t)\,.
\end{eqnarray}
Using the definition of $\breve{k}$ in \eqref{eq-exact-adapt-Volterra-eqn}, as well as the inverse backstepping transformation, $\hat u = w - \hat\beta *w$, we arrive at \eqref{eq:wt_identifier_def}.

\bibliography{references}
\bibliographystyle{plain}

\end{document}